\documentclass[twocolumn, twocolappendix]{aastex631}

\received{April 22, 2022}
\revised{June 02, 2022}
\accepted{June 03, 2022}

\submitjournal{ApJ}

\shorttitle{The \textit{Hubble Space Telescope} UV Legacy Survey of Galactic Globular Clusters. XXIII}

\shortauthors{Libralato et al.}

\usepackage{xspace}
\usepackage{amsmath}
\usepackage{rotating}
\usepackage{threeparttable}
\usepackage{enumerate}

\begin{document}

\title{The \textit{Hubble Space Telescope} UV Legacy Survey of  Galactic Globular Clusters. XXIII. Proper-motion catalogs and internal kinematics.}

\correspondingauthor{Mattia Libralato}
\email{libra@stsci.edu}

\author[0000-0001-9673-7397]{Mattia Libralato}
\affil{AURA for the European Space Agency (ESA), Space Telescope Science Institute, 3700 San Martin Drive, Baltimore, MD 21218, USA}

\author[0000-0003-3858-637X]{Andrea Bellini}
\affil{Space Telescope Science Institute 3700 San Martin Drive, Baltimore, MD 21218, USA}

\author[0000-0003-2742-6872]{Enrico Vesperini}
\affil{Department of Astronomy, Indiana University, Bloomington, IN 47405, USA}

\author[0000-0002-9937-6387]{Giampaolo Piotto}
\affil{Dipartimento di Fisica e Astronomia, Universit\`a di Padova, Vicolo dell'Osservatorio 3, Padova, I-35122, Italy}
\affil{INAF - Osservatorio Astronomico di Padova, Vicolo dell'Osservatorio 5, Padova, I-35122, Italy}

\author[0000-0001-7506-930X]{Antonino P. Milone}
\affil{Dipartimento di Fisica e Astronomia, Universit\`a di Padova, Vicolo dell'Osservatorio 3, Padova, I-35122, Italy}
\affil{INAF - Osservatorio Astronomico di Padova, Vicolo dell'Osservatorio 5, Padova, I-35122, Italy}

\author[0000-0001-7827-7825]{Roeland P. van der Marel}
\affil{Space Telescope Science Institute 3700 San Martin Drive, Baltimore, MD 21218, USA}
\affil{Center for Astrophysical Sciences, The William H. Miller III Department of Physics \& Astronomy, Johns Hopkins University, Baltimore, MD 21218, USA}

\author[0000-0003-2861-3995]{Jay Anderson}
\affil{Space Telescope Science Institute 3700 San Martin Drive, Baltimore, MD 21218, USA}

\author[0000-0002-6054-0004]{Antonio Aparicio}
\affil{Instituto de Astrof\`isica de Canarias, E-38200 La Laguna, Tenerife, Canary Islands, Spain}
\affil{Department of Astrophysics, University of La Laguna, E-38200 La Laguna, Tenerife, Canary Islands, Spain}

\author[0000-0001-9264-4417]{Beatriz Barbuy}
\affil{Universidade de S\~ao Paulo, IAG, Rua do Mat\~ao 1226, Cidade Universit\'asria, S\~ao Paulo 05508-900, Brazil}

\author[0000-0003-4080-6466]{Luigi R. Bedin}
\affil{INAF - Osservatorio Astronomico di Padova, Vicolo dell'Osservatorio 5, Padova, I-35122, Italy}

\author[0000-0003-0066-9268]{Luca Borsato}
\affil{INAF - Osservatorio Astronomico di Padova, Vicolo dell'Osservatorio 5, Padova, I-35122, Italy}

\author[0000-0001-5870-3735]{Santi Cassisi}
\affil{INAF - Osservatorio Astronomico di Abruzzo, Via M. Maggini, s/n, Teramo, I-64100, Italy}
\affil{INFN - Sezione di Pisa, Largo Pontecorvo 3, Pisa, I-56127, Italy}

\author[0000-0003-4237-4601]{Emanuele Dalessandro}
\affil{INAF - Osservatorio di Astrofisica e Scienza dello Spazio di Bologna, Via Gobetti 93/3, Bologna, I-40129, Italy}

\author[0000-0002-2165-8528]{Francesco R. Ferraro}
\affil{Dipartimento di Fisica e Astronomia, Universit\`a of Bologna, Via Gobetti 93/2, Bologna, I-40129, Italy}
\affil{INAF - Osservatorio di Astrofisica e Scienza dello Spazio di Bologna, Via Gobetti 93/3, Bologna, I-40129, Italy}

\author{Ivan R. King}\altaffiliation{We dedicate this paper to the memory of Prof. Ivan R. King, member of this collaboration who passed away on August 31, 2021, whose pioneering work led to a deeper understanding of globular clusters.}
\affil{Department of Astronomy, University of Washington, Box 351580, Seattle, WA 98195, USA}

\author[0000-0001-5613-4938]{Barbara Lanzoni}
\affil{Dipartimento di Fisica e Astronomia, Universit\`a of Bologna, Via Gobetti 93/2, Bologna, I-40129, Italy}
\affil{INAF - Osservatorio di Astrofisica e Scienza dello Spazio di Bologna, Via Gobetti 93/3, Bologna, I-40129, Italy}

\author[0000-0003-1149-3659]{Domenico Nardiello}
\affil{INAF - Osservatorio Astronomico di Padova, Vicolo dell'Osservatorio 5, Padova, I-35122, Italy}

\author[0000-0001-7939-5348]{Sergio Ortolani}
\affil{Dipartimento di Fisica e Astronomia, Universit\`a di Padova, Vicolo dell'Osservatorio 3, Padova, I-35122, Italy}
\affil{INAF - Osservatorio Astronomico di Padova, Vicolo dell'Osservatorio 5, Padova, I-35122, Italy}

\author[0000-0001-6708-4374]{Ata Sarajedini}
\affil{Department of Physics, Florida Atlantic University, Boca Raton, FL 33431, USA}

\author[0000-0001-8368-0221]{Sangmo Tony Sohn}
\affil{Space Telescope Science Institute 3700 San Martin Drive, Baltimore, MD 21218, USA}

\begin{abstract}
  A number of studies based on the data collected by the \textit{Hubble Space Telescope} (\textit{HST}) GO-13297 program ``HST Legacy Survey of Galactic Globular Clusters: Shedding UV Light on Their Populations and Formation” have investigated the photometric properties of a large sample of Galactic globular clusters and revolutionized our understanding of their stellar populations. In this paper, we expand upon previous studies by focusing our attention on the stellar clusters' internal kinematics.  We computed proper motions for stars in 56 globular and one open clusters by combining the GO-13297 images with archival \textit{HST} data. The astro-photometric catalogs released with this paper represent the most complete and homogeneous collection of proper motions of stars in the cores of stellar clusters to date, and expand the information provided by the current (and future) \textit{Gaia} data releases to much fainter stars and into the crowded central regions. We also census the general kinematic properties of stellar clusters by computing the velocity-dispersion and anisotropy radial profiles of their bright members. We study the dependence on concentration and relaxation time, and derive dynamical distances. Finally, we present an in-depth kinematic analysis of the globular cluster NGC~5904.\looseness=-4
\end{abstract}


\section{Introduction}

Over the past 20 yr, photometric and spectroscopic data have radically changed our picture of Galactic globular clusters (GCs). One of the most baffling discoveries is the presence of multiple stellar populations (mPOPs) in GCs \citep[see Anderson 1997;][and references therein]{1999Natur.402...55L, 2004ApJ...605L.125B, 2009A&A...505..139C, 2009A&A...505..117C, 2015AJ....149...91P, 2015MNRAS.454.4197R, 2018ARA&A..56...83B, 2012A&ARv..20...50G, 2019A&ARv..27....8G,2020A&ARv..28....5C}. To shed light on the formation and evolution of mPOPs, the ``HST Legacy Survey of Galactic Globular Clusters: Shedding UV Light on Their Populations and Formation" \citep[GO-13297, PI: Piotto;][]{2015AJ....149...91P} was devised. UltraViolet (UV) and optical data of this and other \textit{Hubble Space Telescope} (\textit{HST}) programs have allowed the creation of color-magnitude (CMDs) and color-(pseudo-)color diagrams, which, in turn, have provided essential elements for better understanding the mPOP phenomenon in Galactic GCs \citep[e.g.,][]{2013MNRAS.431.2126M,2017MNRAS.464.3636M}.\looseness=-4

Other recent important observational findings concern the internal kinematics of GCs. Studies on the internal motions of these systems have revealed that GCs are characterized by complex internal kinematic properties including velocity anisotropy, rotation, and partial energy equipartition \citep[see the review of][and references therein]{2018ComAC...5....2V}. These results have provided the motivation for new theoretical studies aimed at building the theoretical framework necessary to interpret these observational findings \citep[e.g.,][]{2017MNRAS.471.1181B,2018MNRAS.475L..96B,2016MNRAS.461..402T,2017MNRAS.469..683T,2018MNRAS.475L..86T,2019MNRAS.487.5535T,2017MNRAS.471.2778B,2021MNRAS.502.4762B,2019ApJ...887..123S,2021MNRAS.504L..12P,2022MNRAS.509.3815P}. Most of what we have learned on this topic comes from internal motions in the plane of the sky obtained with \textit{HST} and \textit{Gaia} data. These two space observatories have different characteristics that make them best suited for specific, yet complementary, investigations of GCs.\looseness=-4

The \textit{Gaia} mission has revitalized astrometry. The availability of high-precision proper motions (PMs) over the entire sky has enabled a large variety of investigations, for example, large-scale structures in the Galaxy, Galaxy kinematics, stellar streams, tidal tails \citep[e.g.,][]{2018A&A...616A..11G,2019ApJ...872..152I,2021ApJ...914..123I}. The internal kinematics of GCs have benefited from the \textit{Gaia} PMs as well \citep[e.g.,][]{2018MNRAS.481.2125B,2019MNRAS.487.3693J,2021MNRAS.505.5978V,2021arXiv210910998E}, but these \textit{Gaia}-based analyses are focused on (and limited to) bright stars outside the centermost regions. Crowding \citep{2017MNRAS.467..412P} and faintness (i.e., access to low-mass stars) are two hurdles that will be complex (or impossible) to overcome even in the next \textit{Gaia} data releases. However, there is important information in the cores of GCs and in their faint members that is necessary to properly characterize GCs, and one of the few ways of obtaining these necessary data is with \textit{HST}.\looseness=-4

Here we combine the wealth of information available in the \textit{HST} archive to compute high-precision PMs with the goal of analyzing the internal motions within GCs \citep[following][]{2014ApJ...797..115B}. The internal kinematics have a lot to tell, not only about GCs as a whole, but also about their mPOPs. Indeed, the present-day trends in the velocity-dispersion and anisotropy radial profiles are the results of the different initial conditions of first- and second-generation stars. Thus, measuring the internal motions of the mPOPs can help us to shed light on key aspects of the mPOP formation and evolution \citep[e.g.,][]{2013ApJ...779...85M,2016ApJ...823...61M,2015MNRAS.450.1164H,2016MNRAS.455.3693T,2019MNRAS.487.5535T,2013MNRAS.429.1913V,2021MNRAS.502.4290V,2021MNRAS.502.1974S}. This is one of the remaining goals of the GO-13297 program. We present the PM catalogs for the 56 globular and one open clusters targeted by the GO-13297 program, and an overview of the kinematics of their brightest (and more massive) members. To showcase the quality of our catalogs, we have also analyzed in great detail the internal kinematics of NGC~5904.\looseness=-4

Together with the PM catalogs, we release several photometric catalogs that are useful for selecting high-quality objects for studying the internal motions. Recently, \citet{2018MNRAS.481.3382N} published the final version of the photometric catalogs of this project. The photometry presented in this work is not meant to replace that published by \citet{2018MNRAS.481.3382N}. Indeed, although the photometric precision is comparable, the completeness of the catalogs presented here is lower because they contain only sources with positions measured in at least two epochs to enable the determination of the PM. Also, we only release the photometric catalogs of the images in the filters/cameras actually used for the PM computation.\looseness=-4

\section{Data sets and reduction}\label{datared}

One of the goals of this project is to provide high-precision, \textit{homogeneous} PMs from \textit{heterogeneous} \textit{HST} data (i.e. observations with different cameras and programs) for stars in the central regions of GCs. We aimed at computing reliable PMs for as many stars as possible, with particular care for sources in the most crowded regions of the field of view (FoV) and for stars as faint as possible that we can detect with the available \textit{HST} archival images. No two data sets are alike, thus a careful data reduction has been specifically tailored to each GC. Although the data-reduction process is detailed in various papers \citep{2018ApJ...853...86B, 2018ApJ...861...99L, 2019ApJ...873..109L, 2018MNRAS.481.3382N}, we provide here a brief overview and the main differences in our reduction with respect to these works. Two clusters were processed independently as part of other publications: NGC~362 \citep{2018ApJ...861...99L} and NGC~6352 \citep{2019ApJ...873..109L}. For these, we refer the reader to the corresponding papers for detailed descriptions of the data reduction.\looseness=-4

We made use of \texttt{\_flt} images (which are dark and bias corrected, and have been flat-fielded, but not resampled) taken with the Wide-Field Channel (WFC) and High Resolution Camera (HRC) of the Advanced Camera for Surveys (ACS), and with the Ultraviolet-VISible (UVIS) and InfraRed (IR) channels of the Wide-Field Camera 3 (WFC3) taken before 2019\footnote{For NGC~6121, we excluded data from the GO-12911 program, which are part of a more detailed analysis in progress on this cluster.}. In the case of ACS/WFC and WFC3/UVIS data, images were pipeline-corrected for the charge-transfer-efficiency (CTE) defects, as described in \citet{2010PASP..122.1035A}. As discussed in \citet{2014ApJ...797..115B}, not all filters/cameras are suitable for PMs. However, some parts of the reduction process (e.g., the second-pass photometry described below) take advantage of a large number of images to better detect faint sources. For this reason, we chose to include all images at our disposal in the first part of the data reduction\footnote{The data sets used in this work are collected at \protect\dataset[10.17909/gajx-kf45]{http://dx.doi.org/10.17909/gajx-kf45}. All our data products are available at MAST as a High Level Science Product via \protect\dataset[10.17909/jpfd-2m08]{\doi{10.17909/jpfd-2m08}}. See also: \href{https://archive.stsci.edu/hlsp/hacks}{https://archive.stsci.edu/hlsp/hacks}.}.\looseness=-4

Our data reduction is a combination of first- and second-pass photometric stages. First-pass photometry was used to create an initial set of positions and fluxes for the brightest and most isolated sources in each exposure via effective-point-spread-function (ePSF) fitting through a single wave of finding. The ePSFs were specifically tailored to each image, starting from the publicly-available, spatially-variable library of \textit{HST} ePSFs\footnote{\href{https://www.stsci.edu/~jayander/HST1PASS/LIB/}{https://www.stsci.edu/$\sim$jayander/HST1PASS/LIB/}}. Source positions were also corrected for geometric distortion by means of the distortion solutions provided by \citet{2004acs..rept...15A, 2006acs..rept....1A}, \citet{2016wfc..rept...12A}, \citet{2009PASP..121.1419B} and \citet{2011PASP..123..622B}.\looseness=-4

Bright, unsaturated stars in the single-image catalogs were cross-matched with those in the \textit{Gaia} Data Release 2 (DR2) catalog \citep{2016A&A...595A...1G, 2018A&A...616A...1G}. This step was necessary to set up a common, pixel-based reference frame with specific axis orientation (X and Y axes toward west and north, respectively) and pixel scale (40 mas pixel$^{-1}$). The centers of the clusters (from \citealt{2010AJ....140.1830G} with the exception of NGC~5897 and NGC~6791 for which we used the coordinates from the \citealt{1996AJ....112.1487H} catalog, 2010 edition, and Simbad database\footnote{\href{http://simbad.u-strasbg.fr/simbad/}{http://simbad.u-strasbg.fr/simbad/}}, respectively) were placed at a specific position, for example (5000, 5000), so to always have positive master-frame coordinates (the exact coordinates of the center of the clusters is provided in the header of the published catalogs). As in \citet{2018ApJ...853...86B} and \citet{2019ApJ...873..109L}, a master frame was created for every filter, camera and epoch. Then, we iteratively cross-identified the same stars in all images, and applied six-parameter linear transformations to transform the stellar positions in each single-image catalog onto the master frame. Once on the same reference system, the positions and instrumental magnitudes (rescaled to the magnitude of the longest exposure in each epoch/camera/filter) were averaged to create a preliminary astro-photometric catalog.\looseness=-4

The second-pass photometry was obtained with the software \texttt{KS2} \citep[Anderson et al., in preparation; see also][]{2017ApJ...842....6B}. \texttt{KS2} makes use of all images at once to increase the signal of faint objects that would otherwise be undetected in a single image. Starting from the brightest sources, \texttt{KS2} progressively finds fainter stars, and measures their position and flux via ePSF fitting after all detected close-by neighbors have been subtracted from the image. This step is particularly important in crowded environments like the cores of GCs. We run \texttt{KS2} separately for different epochs, by grouping data taken 1--2 yrs apart, to retain stars that have moved by more than one pixel from one epoch to another \citep[see][]{2018ApJ...853...86B}. \texttt{KS2} allows us to define a specific set of images that can be ``stacked" together and used to find sources in the FoV. There is not a one-size-fits-all solution for selecting these reference exposures. For each cluster, we selected a combination of cameras/detectors/filters that provided a good compromise between using a large number of exposures to more easily identify faint stars, and avoiding spurious detections (caused, for example, by the mix of filters with very different wavelength coverages).\looseness=-4

Instrumental magnitudes in the \texttt{KS2}-based catalogs were converted on to the VEGA-mag system. Photometry obtained with ACS/WFC F606W, ACS/WFC F814W, WFC3/UVIS F336W or WFC3/UVIS F438W filters was registered on to the VEGA-mag system by computing the zero-point difference with the corresponding photometry in the catalogs of \citet{2018MNRAS.481.3382N}. The zero-point takes into account for the normalization to 1-s exposure time, the aperture correction and the VEGA-mag zero-point. Photometry with all other cameras/filters was calibrated as described in \citet{2017ApJ...842....6B} by means of \texttt{\_drz} (for ACS/HRC) or \texttt{\_drc} (for ACS/WFC or WFC3/UVIS) images and the official aperture corrections and VEGA-mag zero-points\footnote{See the resources provided here: \href{https://www.stsci.edu/hst/instrumentation/acs/data-analysis}{https://www.stsci.edu/hst/instrumentation/acs/data-analysis}, \href{https://www.stsci.edu/hst/instrumentation/wfc3/data-analysis/photometric-calibration}{https://www.stsci.edu/hst/instrumentation/wfc3/data-analysis/photometric-calibration}.}.

Finally, \texttt{KS2} provides the position and flux of all the detected sources in the raw reference-frame system of each image \citep{2018ApJ...853...86B}. We made use of these \texttt{KS2}-based single catalogs to compute our PMs.\looseness=-4

\begin{figure*}
\centering
\includegraphics[width=\textwidth]{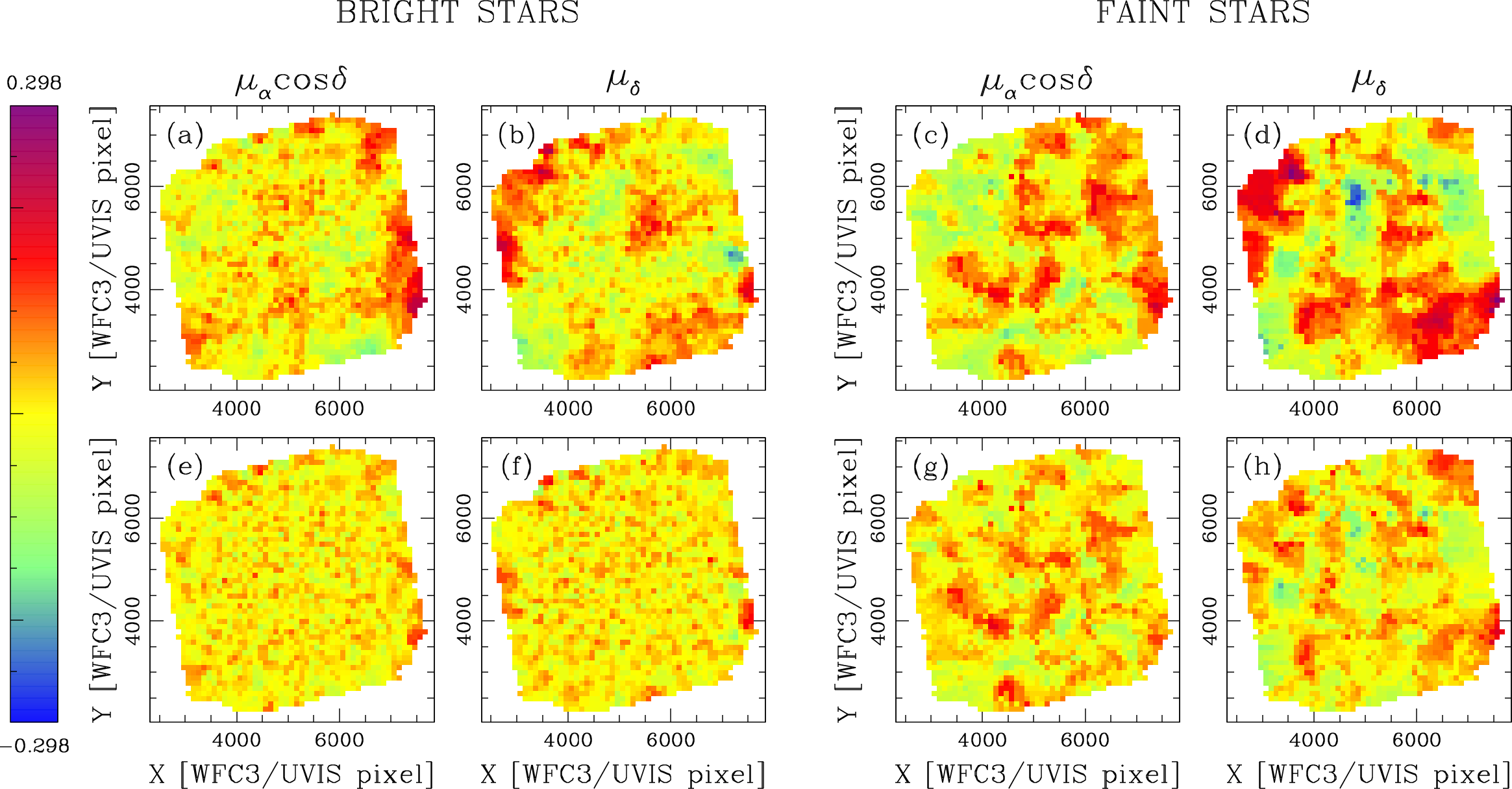}
\caption{Local PM maps of NGC~5272 before (panels a, b, c, d) and after (panels e, f, g, h) the a-posteriori corrections. The four leftmost/rightmost panels are obtained using only stars brighter/fainter than the instrumental F606W magnitude of $-10$ (signal-to-noise ratio of $\sim$100; $m_{\rm F606W} \sim 21.6$). Points are color-coded according to the color bar on the left (in units of mas yr$^{-1}$). See the text for details.}
\label{fig:pmcorrm3}
\end{figure*}

\begin{figure*}
\centering
\includegraphics[width=\textwidth]{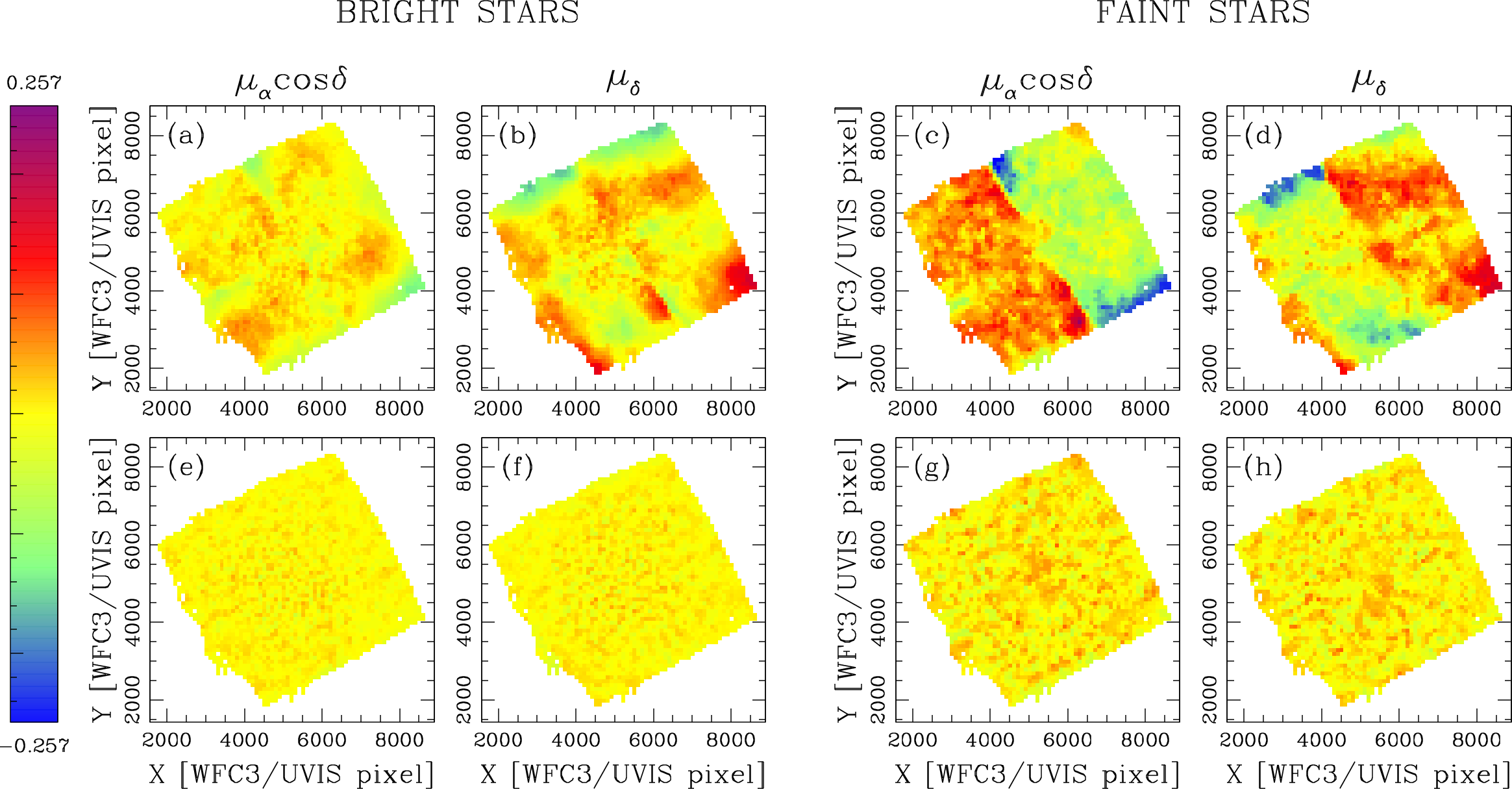}
\caption{Similar to Fig.~\ref{fig:pmcorrm3}, but for the GC NGC~1261. The effects of residual, uncorrected CTE are clearly visible in panels (c) and (d).}
\label{fig:pmcorrngc1261}
\end{figure*}

\section{Proper motions}\label{pms}

PMs were computed following the procedures and caveats described in \citet{2014ApJ...797..115B}. Briefly, positions in the \texttt{KS2}-based raw catalogs were corrected for geometric distortion and then transformed (with six-parameter linear transformations) onto the same reference-frame system defined in Sect.~\ref{datared}. Only cameras/detectors/filters best suited for astrometry\footnote{Besides the filters for which we do not have an ad-hoc geometric-distortion correction, we did not use filters bluer than F336W for the WFC3/UVIS detector and F330W for the ACS/HRC camera, respectively. We also excluded all WFC3/IR data given the worse resolution of the WFC3/IR detector.} \citep[see][]{2014ApJ...797..115B} were used in the PM computation. Then, the positions as a function of time were fit with a least-squares straight line. The slope of this straight line provides a direct estimate of the PM of the source. The PM errors are the uncertainties of the PM fit obtained by using the actual residuals of the points about the fitted line \citep{2014ApJ...797..115B}\footnote{Objects with a peculiar motion that cannot be modeled by a simple straight-line fit, like wide binaries, would result in large PM errors. However, these objects are hard to be discerned from single stars with poorly-measured proper motions from the information in our catalogs alone. A systematic search and accurate PM estimate for objects with very-peculiar motions would require an ad-hoc analysis, which is outside the scope of this project.}.\looseness=-4

For each source, the six-parameter linear transformations used in the PM computation were obtained by using a set of close-by, bright, well-measured cluster stars. Thus, our PMs are computed relative to the bulk motion of each GC at that given specific location in the FoV, and the cluster PM distribution is centered on the origin of the vector-point diagram (VPD). In addition, we also provide the PM zero-point needed to transform these relative PMs on to an absolute reference system (Appendix~\ref{abspm}).\looseness=-4

Another important feature of our PM derivation is that any signature of internal cluster rotation in the plane of the sky is removed from the cluster stars, but it is present, with the opposite sign, in all other sources \citep{2017ApJ...844..167B}. Thus, we cannot directly infer cluster rotation from the kinematics of the cluster members. The same argument is also valid for the parallax effect \citep[e.g.][]{2018ApJ...861...99L,2018ApJ...854...45L}.\looseness=-4

Small spatially-variable and magnitude-dependent systematic errors are present in our uncorrected PMs. As in \citet{2018ApJ...853...86B}, we notice two main types of systematic errors:\looseness=-4
\begin{itemize}
\item a low-frequency effect correlated with the temporal baseline, as well as number and types of images, used to compute the PMs. To remove this time-dependent systematic, we divided our sample in $N$ sub-groups based on the PM temporal baseline. $N$ varies from cluster to cluster due to the heterogeneous data sets used. Then, we computed the median PM of each sub-group, which should be zero by construction. If it was not zero, we subtracted this median PM value to the PM of each star in the sub-group;\looseness=-4
\item a high-frequency, spatially- and magnitude-dependent systematic error. By construction, the average PM of cluster stars should be zero regardless their magnitude and location in the field. This is not always true locally, mainly because of a combination of CTE and geometric-distortion residuals. These residual systematic errors were corrected using the median PM of the closest, both spatially and in magnitude, $N$ well-measured cluster members (target star excluded). The closeness criterion and the number of reference stars $N$ were tailored to each cluster to reach the best compromise between mapping the variations as locally as possible and the need for large statistics. For very-bright(faint) objects, we set up a magnitude threshold above(below) which reference stars are used for the correction instead of a fixed $\Delta$mag. This was done to increase the statistics at the extreme ends of the magnitude range.\looseness=-4
\end{itemize}
The errors that we report in our catalogs for the PMs thus corrected include the propagated contributions from the uncertainties in the corrections themselves. When no enough reference stars were available for the high-frequency correction, this correction was not applied.\looseness=-4

Figure~\ref{fig:pmcorrm3} shows maps of raw and a-posteriori, locally-corrected PMs for the GC NGC~5272. The FoV has been divided into square cells of 100 WFC3/UVIS pixels per side. In each cell, we selected the 50 well-measured, cluster stars closest to the center of the cell, and computed the average PM in each direction. Panels (a) to (d) present the local PM maps obtained by means of the raw, uncorrected PMs, while panels (e) to (h) show the corrected PM maps. In each row, the two leftmost panels refer to stars brighter than the F606W instrumental magnitude equal to $-10$ (signal-to-noise ratio of $\sim$100; $m_{\rm F606W} \sim 21.6$), while the maps for fainter stars are shown in the two rightmost panels. The comparison between the top and bottom panels clearly highlights the effectiveness of the a-posteriori corrections. However, some residual high-frequency systematic errors are still present among the faintest objects. Thus, caution is advised when using these objects.\looseness=-4

The raw PMs of some clusters, mainly those including the ACS/WFC data of the GO-14235 program (PI: Sohn), present larger systematics related to uncorrected CTE. The CTE that affects the \textit{HST} detectors has worsened over time, and the official pipeline is not always able to completely correct it. An example is shown for NGC~1261 in the top panels of Figure~\ref{fig:pmcorrngc1261}, where it is clear that (i) bright and faint stars have different systematic errors in the PMs, and (ii) the local PM map of faint stars presents a discontinuity in the FoV along the chip gap of the GO-14235 data set. For these specific cases, we applied an additional a-posteriori correction prior to that for high-frequency systematics. Briefly, we divided our sample of well-measured stars into four magnitude bins. In each sub-sample, we computed the average PM (in each direction) of cluster members in 125-pixel-wide bins along the direction perpendicular to that of the CTE systematic on the local PM map. The corrections to the PM of each star were computed by interpolating among these binned values. As for the other a-posteriori corrections, the errors of this CTE-related correction are included in the corrected-PM error budget.\looseness=-4

\begin{figure*}[t!]
\centering
\includegraphics[width=\textwidth]{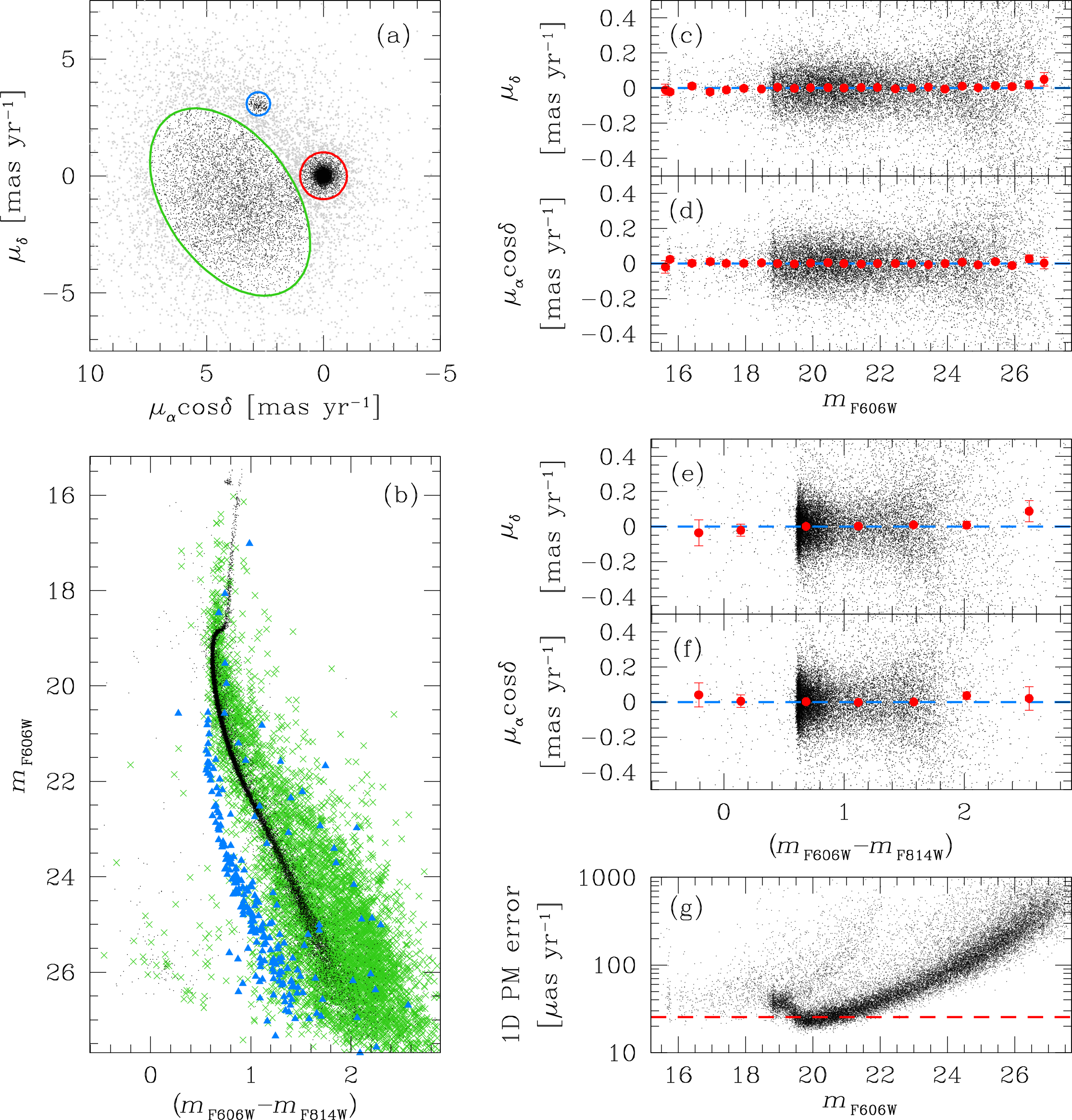}
\caption{Overview of the PM catalog of NGC~6652. (a): VPD of the relative, corrected PMs. Stars within the red circle (centered on the origin of the VPD, and with radius of 1 mas yr$^{-1}$) are likely cluster members, the blue circle marks the location of Sagittarius-Dwarf objects, while the green ellipse highlights a group of Bulge stars. Gray dots are all sources outside any of these three selections (likely Bulge objects or cluster stars with large PM uncertainties). (b): $m_{\rm F606W}$ versus $(m_{\rm F606W}-m_{\rm F814W})$ CMD. Black dots are likely cluster members; blue triangles are members of the Sagittarius Dwarf; green crosses are Bulge objects (gray points in the VPD are not shown for clarity). (c) and (d): $\mu_\delta$ and $\mu_\alpha \cos\delta$ PMs as a function of $m_{\rm F606W}$. Only stars member of NGC~6652 are shown (black points). The red points (with error bars) are the median values of the PMs in 0.5-mag bins. The azure line is set to zero. (e) and (f): $\mu_\delta$ and $\mu_\alpha \cos\delta$ PMs as a function of $(m_{\rm F606W}-m_{\rm F814W})$. Panels (c) to (f) show no obvious trends of our corrected PMs with stellar magnitude and color. (g) 1D corrected-PM error (the sum in quadrature of the PM errors in each direction divided by $\sqrt{2}$) as a function of $m_{\rm F606W}$. The red line is set at the median value of the 1D PM errors of bright, well-measured unsaturated stars.}
\label{fig:overview}
\end{figure*}

The steps described above generally remove the majority of systematic errors included in the raw PMs. However, these corrections are not perfect, especially for very faint stars, and we advise users to carefully check PMs for magnitude/color/spatial systematics on a cluster-by-cluster basis.\looseness=-4

\begin{figure*}
\centering
\includegraphics[width=\columnwidth]{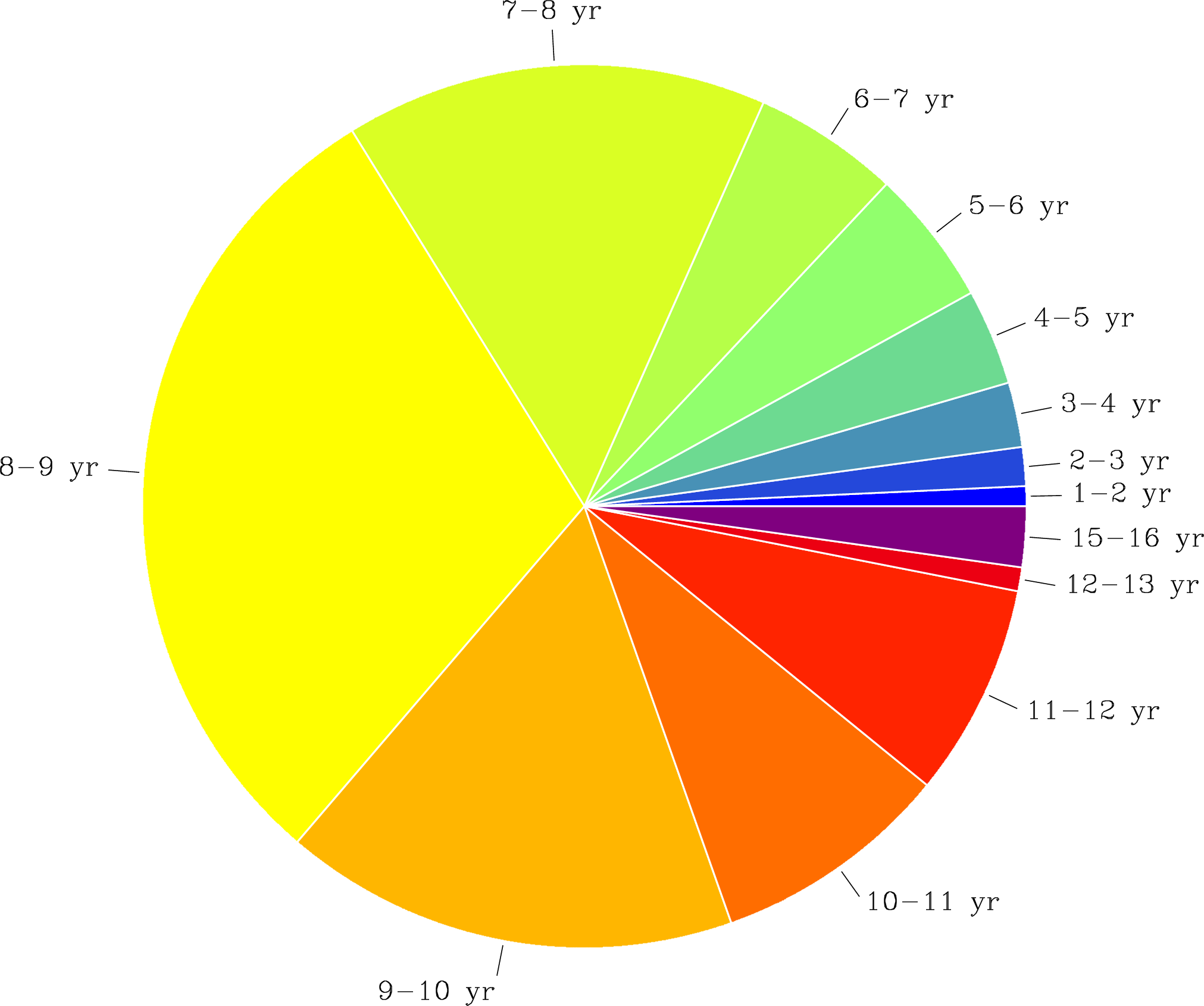}
\includegraphics[width=\columnwidth]{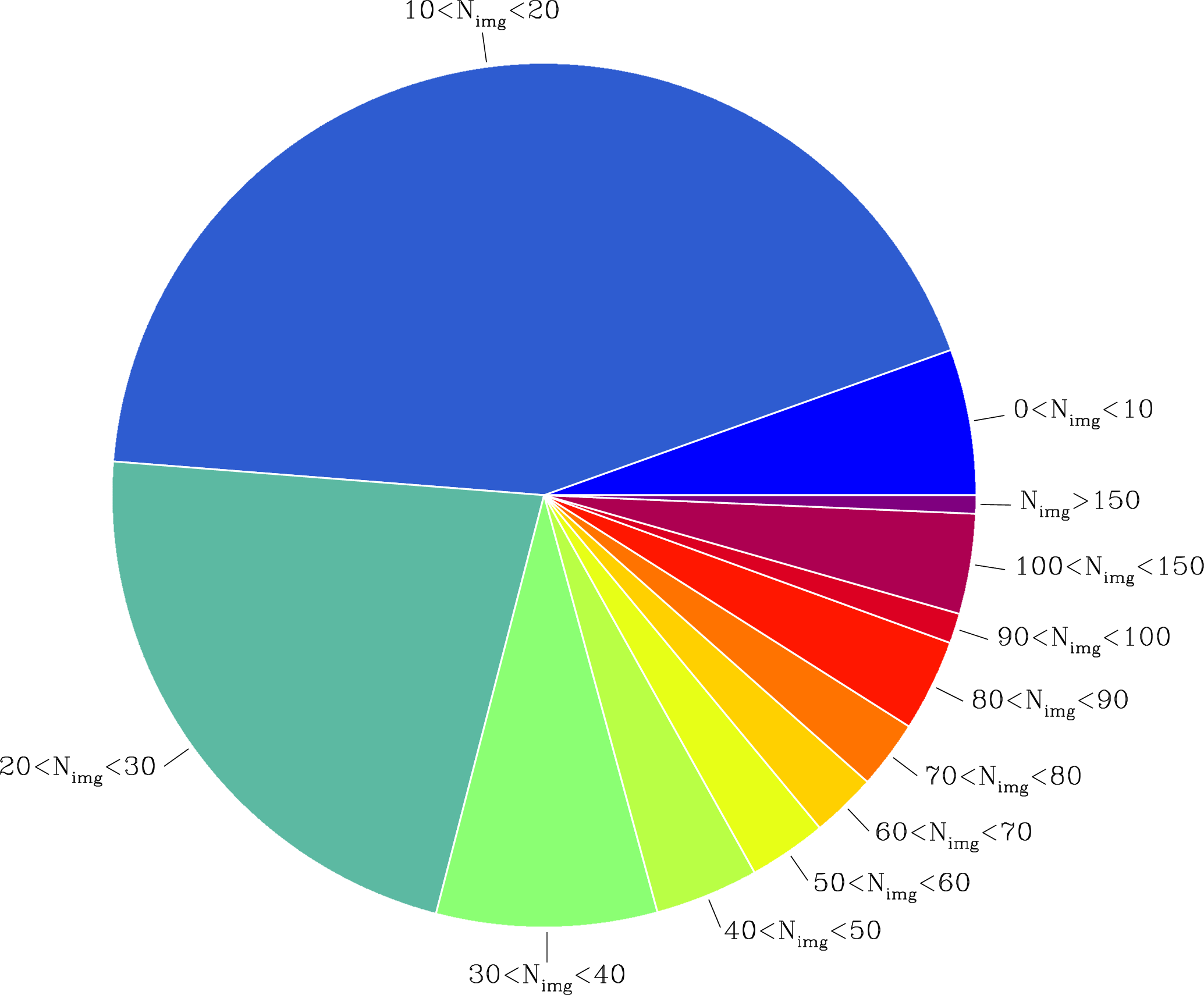}
\caption{Pie charts of the total temporal baselines (left) and number of images used to compute PMs (right) in our catalogs.}
\label{fig:pies}
\end{figure*}

Figure~\ref{fig:overview} provides an overview of the PM catalog of NGC~6652. This GC is located in the outer Bulge, projected on the Baade's Window \citep{2015MNRAS.450.3270R}, and in foreground of the Sagittarius Dwarf spheroidal. The VPD obtained with our corrected PMs is presented in panel (a). On the basis of their locations in the VPD, we arbitrarily define three groups of objects: cluster stars (points within the red circle), Sagittarius-Dwarf members (points within the blue circle), and Bulge stars (objects within the green ellipse). In the CMD in panel (b), we highlight members of NGC~6652 in black (stars within the red circle in the VPD), Bulge objects in green and stars associated with the Sagittarius Dwarf in azure. This is a simple example of one of the possible applications of our PM catalogs. Corrected PMs in each coordinate as a function of $m_{\rm F606W}$ are plotted in panels (c) and (d), and as a function of $(m_{\rm F606W}-m_{\rm F814W})$ color in panels (e) and (f). Only cluster members are shown. The red points (with error bars) are the median values of the PMs in 0.5-mag bins. The azure line is set to zero. These plots do not show any significant magnitude- or color-dependent systematics in our corrected PMs, but very blue and red objects hint at the presence of some residual color-dependent systematics. However, stars with these extreme colors are very faint (see panel b), and we expect both PMs and their corrections still to be affected by small systematic residuals. We stress, though, that no quality selections other than the membership were applied here. Finally, panel (g) shows the 1D corrected-PM error as a function of $m_{\rm F606W}$. The red, horizontal line is set at the median 1D PM error of bright, well-measured stars, i.e., 25.6 $\mu$as yr$^{-1}$.\looseness=-4

A description of the final PM and photometric catalogs (together with some caveats about their usage) is provided in Appendix~\ref{cats}, Tables~\ref{tab:pmcat} and \ref{tab:photcat}, respectively. The precision of our PMs varies from cluster to cluster and depends on temporal baseline, as well as on the number and depth of images. Figure~\ref{fig:pies} presents pie charts of the temporal baselines and the number of images used to compute PMs in all our catalogs, respectively. About $\sim$33\% of the PMs are computed with only 10--20 images and a temporal baseline between 7 and 9 yr, i.e., by combining only GO-10775 and GO-13297 data. The remaining PMs result from the mix of heterogeneous data sets. For bright, well-measured stars, the median raw-PM precision is between about 7 and 60 $\mu$as yr$^{-1}$, while that of corrected PMs ranges between about 13 and 120 $\mu$as yr$^{-1}$.\looseness=-4

\section{Internal kinematics of stellar clusters}\label{kin}

As a benchmark for our PM catalogs, we here provide velocity-dispersion and anisotropy radial profiles of the clusters in our project. The kinematic profiles of the cores complement those in the outskirts derived with the Gaia Early Data Release 3 (EDR3) catalog \citep[e.g.,][]{2021MNRAS.505.5978V}. We also include the kinematic profile of the open cluster NGC~6791.\looseness=-4

We started by selecting well-measured objects in each epoch/filter/camera combination as follows (see Table~\ref{tab:photcat} for the explanation of the parameters):
\begin{enumerate}[(i)]
    \item magnitude rms lower than the 90-th percentile of the distribution at any given magnitude. An object with a magnitude rms better than 0.01 mag is always included in the well-measured sample, while all sources with a magnitude rms larger than 0.15 mag are excluded. 
    \item \texttt{QFIT} value larger than the 90-th percentile of the distribution at any given magnitude (note that the closer to 1 the \texttt{QFIT} parameter is, the better the PSF fit). Again, we retained all objects with a \texttt{QFIT} larger than 0.99 and discarded those with a value lower than 0.75.
    \item $|$\texttt{RADXS}$|$ value lower than the 90-th percentile of the distribution at any given magnitude, but keeping all objects with a $|$\texttt{RADXS}$|$ lower than 0.01 and rejecting those with a value larger than 0.15;
    \item the photometric $N_{\rm u}^{\rm phot}/N_{\rm f}^{\rm phot}$ ratio (see Table~\ref{tab:photcat}) is greater than 0.75;
    \item $o < 1$;
    \item flux at least 3.5$\sigma$ above the local sky.
\end{enumerate}
These (empirically-derived) thresholds were chosen as a compromise between rejecting bad measurements and keeping a large sample of stars for the subsequent analyses. To avoid crowding bias, we measured the 90-th-percentile trends for magnitude rms, \texttt{QFIT} and \texttt{RADXS} in a region outside the core of each cluster where sources are more isolated and applied these cuts to all stars across the FoV. When not enough stars were available outside the core of the cluster (for example, for the photometry obtained with ACS/HRC full-frame or WFC3/UVIS subarray images), all stars were used, regardless of their location in the FoV.\looseness=-4

For each epoch, all criteria from (i) to (vi) have to be fulfilled in at least two filters if a star has been measured in at least two filters, otherwise in the only filter through which it was been detected. Finally, an object that has passed all previous selections is defined as ``well-measured" if it passes all criteria in at least two epochs.\looseness=-4

The PMs also provide useful parameters for selecting trustworthy sources for the kinematic analysis. In addition to all the previous criteria, we also removed all objects that have an astrometric $N_{\rm u}^{\rm PM}/N_{\rm f}^{\rm PM}$ ratio (see Table~\ref{tab:pmcat}) smaller than 0.8--0.9 (the exact value changes from cluster to cluster), $\chi^2_{\mu_\alpha \cos\delta}$ and $\chi^2_{\mu_\delta}$ larger than 1.25--1.5, PM error larger than 0.5 mas yr$^{-1}$, or for which the a-posteriori PM correction was not computed. Finally, we also excluded stars with a PM error larger than $f$ times the local velocity dispersion $\sigma_\mu$ of a sub-sample of well-measured, close-by cluster stars. For each cluster, we compared the velocity-dispersion radial profiles of RGB stars obtained by varying $f$ from 0.5 to 0.9, with steps of 0.1, finding a general good agreement (within 1$\sigma$) between the inferred kinematics. If enough stars were left after all our quality selections, we used $f = 0.5$. For 9 GCs (NGC~5053, NGC~5466, NGC~5897, NGC~6144, NGC~6366, NGC~6496, NGC~6535, NGC~6584 and NGC~6717), we used a value of $f = 0.8$, while for NGC~6981 we set $f = 0.9$. We included these 10 clusters in our analysis, but because their PM errors are of the order of their $\sigma_\mu$, we advise caution in the interpretation of their velocity-dispersion profiles.\looseness=-4

All these criteria are designed to obtain a good compromise between statistics, quality and completeness. The only two exceptions to the strategy described above are NGC~362 and NGC~6352, for which we adopted the quality selections described in \citet{2018ApJ...861...99L} and \citet{2019ApJ...873..109L}, respectively.\looseness=-4

GCs are old, collisional systems that, after many two-body relaxation times, present a (partial) degree of energy equipartition. Because of the heterogeneous mass ranges covered by our PM catalogs, we restricted the analysis of the kinematic profiles to stars brighter than the MS turnoff along the sub-giant and red-giant branches (SGB and RGB, respectively). Since in the post-MS evolutionary stages the evolutionary lifetimes are quite shorter than the core H-burning one, and in any case shorter than the typical two-body relaxation time, SGB and RGB stars can be safely considered as having the same kinematic mass (see also Sect.~\ref{weight}).\looseness=-4

The velocity-dispersion profile of each cluster was obtained by dividing the sample of massive cluster members\footnote{The cluster membership was inferred by means of CMDs and PMs. The PM threshold for each cluster was defined as the best compromise between including genuine members with large observed dispersion (which includes the contribution of both the PM errors and intrinsic velocity dispersion), and removing field objects with PMs similar to that of the cluster.} into equally-populated radial bins, with at least 100 stars per bin. In case of low statistics, we lowered the number of stars per bin to ensure at least 3 radial bins, regardless of the number of stars within each bin. When the statistics allowed, we also imposed a radial bin comprising only the centermost 5--10 arcsec. The velocity dispersion in each radial bin was computed similarly to what is described in \citet{2020ApJ...895...15R}, i.e., by maximizing the following likelihood:
\begin{equation}\label{eq:vdisp}
    \begin{aligned}
        \ln \mathcal{L} = -\frac{1}{2} \sum_{n} \Bigl[&\frac{(v_{\rm rad,\,n}-v_{\rm rad})^2}{\sigma_{\rm rad}^2+\epsilon_{\rm rad,\,n}^2} + \ln(\sigma_{\rm rad}^2 + \epsilon_{\rm rad,\,n}^2) + \\
        & \frac{(v_{\rm tan,\,n}-v_{\rm tan})^2}{\sigma_{\rm tan}^2+\epsilon_{\rm tan,\,n}^2} + \ln(\sigma_{\rm tan}^2 + \epsilon_{\rm tan,\,n}^2) \Bigr] \mathrm{ ,}
    \end{aligned}
\end{equation}
where $(v_{\rm rad,\,n}, v_{\rm tan,\,n})$ are the radial and tangential components of the PM of the $n$-th star, $(\epsilon_{\rm rad,\,n}, \epsilon_{\rm tan,\,n})$ are the radial and tangential components of the PM uncertainty of the $n$-th star, $(v_{\rm rad}, v_{\rm tan})$ are the radial and tangential mean motions of the cluster, and $(\sigma_{\rm rad}, \sigma_{\rm tan})$ are the radial and tangential velocity dispersions of the cluster. We also computed the combined velocity dispersion $\sigma_\mu$ in each radial bin using the same likelihood in Eq.~\ref{eq:vdisp} but with $\sigma_{\rm rad}=\sigma_{\rm tan}=\sigma_\mu$. We used the affine-invariant Markov Chain Monte Carlo (MCMC) method \texttt{emcee} \citep{2013PASP..125..306F} to sample the parameter space and obtain the posterior probability distribution functions (PDFs) for $\sigma_\mu$, $\sigma_{\rm rad}$ and $\sigma_{\rm tan}$. We run the MCMC chain with 20 walkers for 5000 steps, then rejected the first 200 steps. The best-fit values correspond to the medians of the PDFs, while the corresponding errors are defined as the average between the 16-th and 84-th percentile about the median. Finally, the velocity dispersions were corrected as described in \citet[but see also Appendix~A of \citealt{2006A&A...445..513V}]{2015ApJ...803...29W} to take into account the maximum-likelihood estimators being biased and underestimating the true dispersion of the velocity distribution. The difference between the corrected and uncorrected $\sigma_\mu$ is, on average, $\sim$0.6\%, it never reaches 3\% and it is more important for bins with fewer than 100 stars.\looseness=-4

Figures~\ref{fig:kin1}--\ref{fig:kin4} in Appendix~\ref{app:kin} show the result of our analysis for the 57 stellar clusters in the GO-13297 project. These profiles are also available at our website\footnote{\href{https://archive.stsci.edu/hlsp/hacks}{https://archive.stsci.edu/hlsp/hacks}}. For each cluster, the $m_{\rm F606W}$ versus $(m_{\rm F606W}-m_{\rm F814W})$ CMD is shown on the rightmost panel. Well-measured members of the GC brighter than the MS turnoff (highlighted by the azure, dashed horizontal line) are plotted as red points, all other objects that passed the quality criteria are shown as black points.\looseness=-4

\begin{table}[t!]
  \caption{\textit{Gaia}-based velocity dispersions for NGC~6791.}
  \centering
  \label{tab:ngc6791}
  \begin{tabular}{cccc}
    \hline
    \hline
    Radius & $\sigma_\mu$ \\
    $[$arcsec$]$ & $[$mas yr$^{-1}$$]$ \\
    \hline
     73.55  & $0.096 \pm 0.005$ \\
     144.58 & $0.074 \pm 0.005$ \\
     222.38 & $0.075 \pm 0.005$ \\
     311.43 & $0.075 \pm 0.005$ \\
     487.58 & $0.066 \pm 0.005$ \\
    \hline
  \end{tabular}
\end{table}

The velocity dispersion $\sigma_\mu$ as a function of distance from the center of the cluster is presented in the bottom-left panel. Black, filled points are obtained from this work, while black, open points refer to the measurements in the GC database of Holger Baumgardt\footnote{\href{https://people.smp.uq.edu.au/HolgerBaumgardt/globular/}{https://people.smp.uq.edu.au/HolgerBaumgardt/globular/}} obtained with the Gaia EDR3 PMs, which were presented by \citet{2021MNRAS.505.5978V}. The only exception is NGC~6791, which is not included in the work of \citet{2021MNRAS.505.5978V}. For this cluster, we independently derived \textit{Gaia}-based $\sigma_\mu$ using our tools\footnote{We considered only cluster stars (selected by means of PMs and CMD) within 750 arcsec from the center of NGC~6791. We rejected all sources that had a re-normalized unit weight error (\texttt{RUWE}) greater than 1.4, an astrometric excess noise larger than 0.4, a number of bad along-scan observations exceeding 1.5\% of the total number of along-scan observations, or a 2D PM error worse than 0.3 mas yr$^{-1}$. We also excluded all objects fainter than the MS turnoff or brighter than $G = 13$.} (Table~\ref{tab:ngc6791}). The black dashed lines are set at the core ($r_{\rm c}$; obtained using the definition in \citealt{1987degc.book.....S}, see Eq.~1-34, which is similar to the definition of the King scale radius $r_0$) and the projected half-light ($r_{\rm h}$) radii provided in the Baumgardt's database. For NGC~6791, we used $r_{\rm c}$ and $r_{\rm h}$ from \citet{2015MNRAS.449.1811D} and \citet{2019MNRAS.483.2197K}, respectively.\looseness=-4

\begin{figure*}[t!]
\centering
\includegraphics[width=\textwidth]{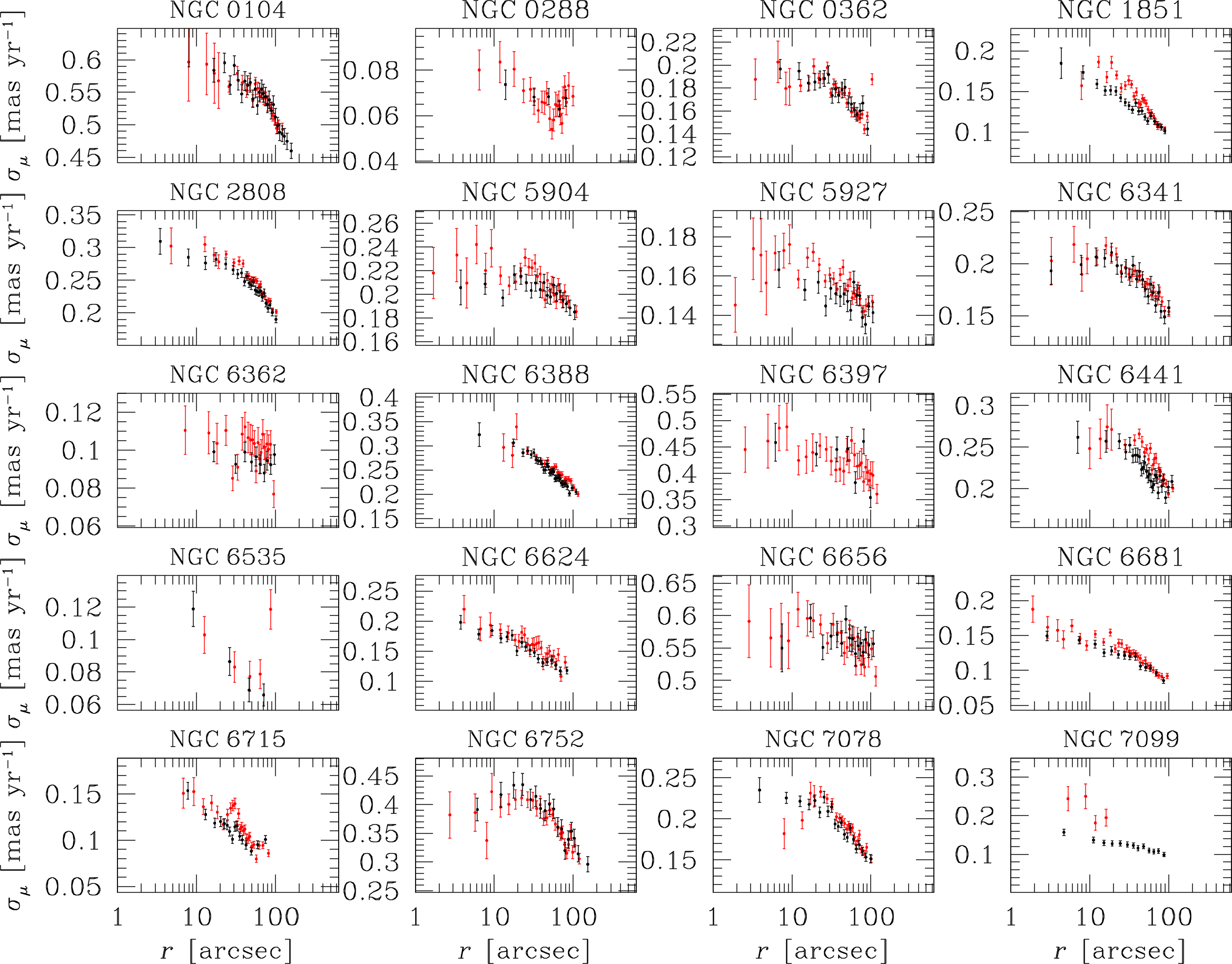}
\caption{Comparison of the velocity-dispersion radial profiles obtained with \textit{HST}-based PMs in this paper (black points) and in \citet[red points]{2015ApJ...803...29W}.}
\label{fig:comp_watkins}
\end{figure*}

\citet{2021MNRAS.505.5978V} used all stars at their disposal regardless of their magnitude. Thus, the \textit{Gaia}-based velocity dispersions for close-by clusters were derived from stars with various masses and, because of the effects of energy equipartition, could be systematically higher than our \textit{HST}-based profiles. However, \citet{2021MNRAS.505.5978V} argued that the \textit{Gaia} uncertainties for faint stars are likely underestimated, and included a scaling factor for the PM errors to obtain consistent $\sigma_\mu$ values between different magnitude intervals. Furthermore, the authors applied various quality cuts to their samples prior to the determination of the PM velocity dispersions (see their Sect.~3), which likely excluded from the fits faint (and so low-mass) objects with large PM uncertainties. For this reason, we choose to directly compare our \textit{HST} profiles in the cores with these \textit{Gaia} velocity dispersions outside the cores. Overall, we find a good agreement between \textit{HST} and \textit{Gaia} profiles, with a few exceptions. NGC~5466 and NGC~6981 have \textit{HST} PM uncertainties of the same order of the $\sigma_\mu$ (we used a large value of $f$ for the analysis), thus the inferred values of $\sigma_\mu$ should be interpreted cautiously. For NGC~6934, most of the \textit{Gaia} stars used by \citet{2021MNRAS.505.5978V} have PM errors larger than the intrinsic $\sigma_\mu$ of the cluster, and the \textit{Gaia} measurements might not be completely reliable. The \textit{HST} profile of NGC~6304 seems higher than the one from the \textit{Gaia} PMs. However, the \textit{HST}-based kinematics are in agreement with the line-of-sight (LOS) velocities in the Baumgardt's database.\looseness=-4

We fit these points with a 4-th order monotonically-decreasing polynomial. The coefficients of the 4-th-order polynomial are obtained with a maximum-likelihood approach. For each cluster in Figures~\ref{fig:kin1}--\ref{fig:kin4}, the blue line in the bottom-left panel is obtained with the best-fit (median) values of the polynomial fit, while the cyan lines are 100 random solutions of the polynomial fit. We used these polynomial functions to derive the $\sigma_\mu$ at the center of each cluster, at $r_{\rm c}$ and at $r_{\rm h}$. We provide all these values in Table~\ref{tab:vdisp}.\looseness=-4

The polynomial fits show again the agreement between \textit{HST} and \textit{Gaia} data for most of the clusters. There are a few points that are outliers with respect to the polynomial-fit predictions. Most of these outliers refer to the velocity dispersion in the innermost bins. A rise of the velocity dispersion in the innermost region can be a proxy for the presence of an intermediate-mass black hole \citep[and references therein]{2020ARA&A..58..257G}, but also for issues related to crowding/blending. Indeed, if two sources are blended and confused as one, or if the light contamination from the neighbors is high (which can still occour even with our data reduction), the position measured can be shifted from the real position. The offset is different for every image/filter/camera. Hence, the net result is that its PM is likely larger than what it should be, thus increasing the $\sigma_\mu$ of the sources in the very-crowded region \citep[see the discussion in][]{2014ApJ...797..115B}. Finally, the shape and the abrupt drop of the polynomial functions in the outermost parts of the FoV do not correspond to physical effects and are just plotted for completeness.\looseness=-4

The anisotropy ($\sigma_{\rm tan}/\sigma_{\rm rad}$) as a function of distance from the center of the cluster is presented in the top-left panels. Anisotropy is discussed in \citet{2021MNRAS.505.5978V}, but their values are not publicly available. Thus, only \textit{HST}-based data are shown in the top-left panels of Figs.~\ref{fig:kin1}--\ref{fig:kin4} (black, filled dots). The red, dashed, horizontal line is set to 1 and marks the isotropic case. Most clusters are isotropic in the core, but a few objects show a radial anisotropy outside about one $r_{\rm h}$. We will discuss the kinematic anisotropy in detail in the next Section.\looseness=-4

\citet{2019MNRAS.487.3693J} analyzed 10 GCs using \textit{Gaia} DR2 data and made publicly available their velocity dispersion and anisotropy radial profiles. Although their profiles do not cover the centermost region observed by our \textit{HST} data, we find an agreement at the 1$\sigma$ level between their $\sigma_\mu$ and anisotropy values for the nine clusters in common with our data set.\looseness=-4

We also compared our kinematic profiles with those of \citet{2015ApJ...803...29W}, see Fig.~\ref{fig:comp_watkins}, where a previous version of the \textit{HST} PM catalogs was used \citep{2014ApJ...797..115B}\footnote{The main differences between the PMs computed in this work and those in \citet{2014ApJ...797..115B} are the following: (i) we have more data and longer temporal baselines at disposal; (ii) we now include a second-pass photometry stage to improve the measurements for faint stars and in crowded environments; and (iii) the PM errors now include the contribution of the a-posteriori systematic corrections.}. There is an overall agreement between the profiles at the 3$\sigma$ level. The profiles of \citet{2015ApJ...803...29W} for NGC~1851, NGC~6441 and NGC~7078 are generally higher than those in our work. For NGC~7078, we also do not see the drop of the velocity dispersion in the centermost region as seen by \citet{2015ApJ...803...29W}. A similar discrepancy is found when comparing the profiles of NGC~2808, NGC~6681 and NGC~6715, although at to lesser extent. The reason for these discrepancies might be the better treatment of crowding in our data reduction and/or the additional quality selections applied in our work. The profile of NGC~7099 is instead completely different, but the PMs used by \citet{2015ApJ...803...29W} were computed with a temporal baseline of only two years \citep[see][]{2014ApJ...797..115B}, and their quality is worse than that of our PMs.\looseness=-4

The velocity-dispersion profile of NGC~6441 was also studied by \citet{2021MNRAS.503.1490H}. The PMs computed by \citet{2021MNRAS.503.1490H} are obtained from a combination of data taken with ACS/HRC@\textit{HST} and NACO@VLT detectors, which are better suited for probing the centermost arcsec of this very-crowded cluster. Our PM profile seems to suggest a more moderate increase of the velocity dispersion toward the center, although the values in the region in common within 10 arcsec from the center are in agreement at the 1$\sigma$ level. The central velocity dispersion inferred in our work with the polynomial fit is $\sigma_\mu^{r=0} = (0.285 \pm 0.012)$ mas yr$^{-1}$, while their $\sigma_\mu$ measured at $r \sim 0.76$ arcsec is $(0.316 \pm 0.034)$ mas yr$^{-1}$. It is hard to clarify the nature of the discrepancy between these profiles given the different resolutions of the instruments used to infer the PMs in the centermost arcsec of the cluster.\looseness=-4

Finally, we compared our PM velocity dispersions with those based on LOS velocities in the Baumgardt database, which are taken from various sources in the literature. There is an overall agreement between the PM- and LOS-velocity-based profiles, although there are discrepancies in some cases (NGC~5053, NGC~5272, NGC~5286, NGC~5466, NGC~6093, NGC~6101, NGC~6144, NGC~6205, NGC~6584, NGC~6681, NGC~6809, NGC~6981 and NGC~7099) where our PM-based $\sigma_\mu$ are higher than the $\sigma_{\rm LOS}$, either in general or only in the centermost bins. The origin of the differences between PM- and LOS-velocity-based profiles might be related to systematics in either data sets, errors in the cluster distance (see Sect.~\ref{kdist}), or instead be a proxy of a peculiar kinematic state of the cluster. However, a detailed comparison between these $\sigma_\mu$ is outside the scope of this paper.\looseness=-4

\section{General kinematic properties}

The collection of kinematic profiles in Figures~\ref{fig:kin1}--\ref{fig:kin4} allows us to analyze some of the general properties of stellar clusters, similar to what has already been done in the literature \citep[e.g.,][]{2015ApJ...803...29W}, but with a larger sample. We cross-correlated the kinematic pieces of information derived by means of \textit{HST} and \textit{Gaia} data with the structural properties of the GCs.\looseness=-4

\begin{figure}[t!]
\centering
\includegraphics[width=\columnwidth]{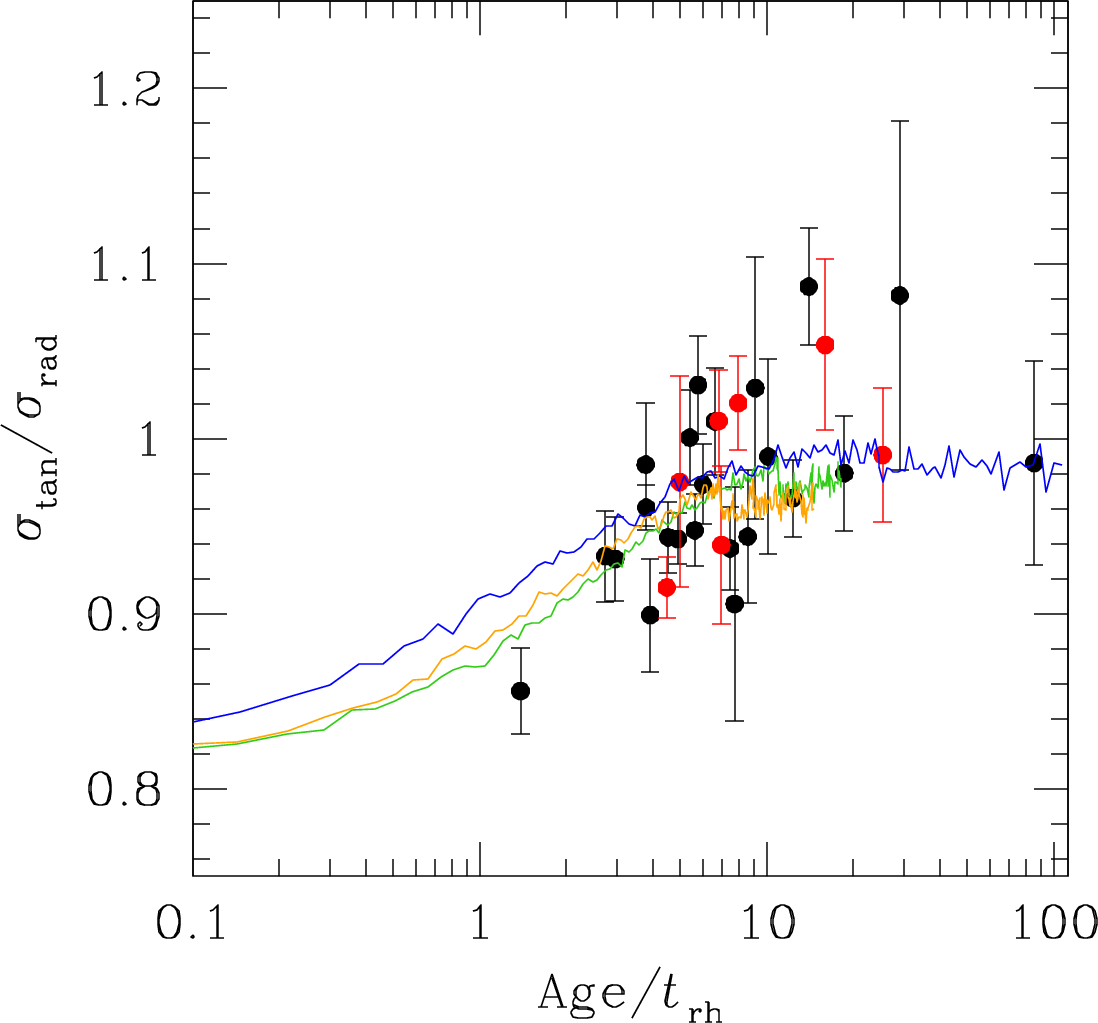}
\caption{Anisotropy at the half-light radius as a function of the ratio of the cluster age to the half-mass relaxation time. An isotropic system is characterized by $\sigma_{\rm tan}/\sigma_{\rm rad} = 1$. Red points refer to core-collapsed GCs. All other systems are shown as black dots. The three solid lines show the time evolution of $\sigma_{\rm tan}/\sigma_{\rm rad}$ from three Monte Carlo simulations with initial $W_0$ and filling factors (defined as the ratio of the half-mass to tidal radius) equal to (5, 0.09; green line), (5, 0.18; blue line), (7, 0.06; orange line); the three systems reach core collapse, respectively, at Age$/t_{\rm rh} \simeq$ 10.9, 24.7, 6.8.}
\label{fig:trel}
\end{figure}

A quantity of particular interest in the kinematic characterization of star clusters is the anisotropy of the velocity distribution. Simulations following the evolution of star clusters during the initial violent-relaxation phase and including the effects of the tidal field of the host galaxy have shown that these systems emerge from this early evolutionary phase with an isotropic velocity distribution in the core, a radially anisotropic distribution in the intermediate regions, and an isotropic or slightly tangentially anisotropic distribution in the outermost regions \citep[see][]{2014MNRAS.443L..79V}. This specific configuration depends on the initial conditions at the formation of the cluster, on the surrounding environment, and it changes during the subsequent long-term evolution; in particular during a cluster's long-term evolution, the effects of two-body relaxation and mass loss lead to a gradual decrease of the radial anisotropy imprinted during the early dynamical phases \citep[][see that paper also for the possible development of radial anisotropy in tidally underfilling clusters with an initial isotropic velocity distribution]{2016MNRAS.455.3693T}.\looseness=-4

\begin{figure*}[th!]
\centering
\includegraphics[width=\textwidth]{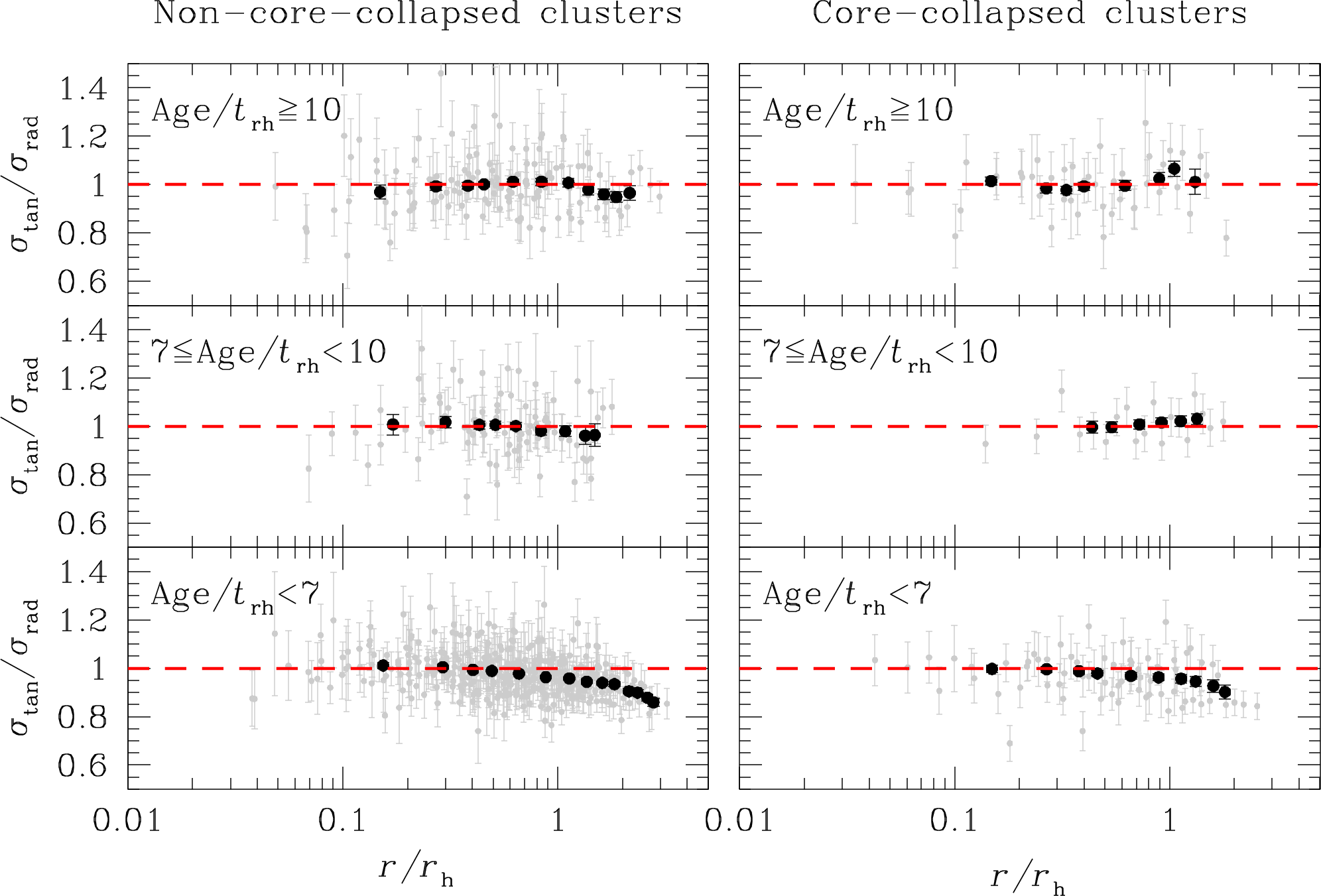}
\caption{Anisotropy as function of distance from the center of the cluster in units of $r_{\rm h}$. The red, dashed, horizontal line is set at 1. Gray dots refer to the individual measurements shown in Figs.~\ref{fig:kin1}--\ref{fig:kin4}. Black points, with error bars, are the 3.5$\sigma$-clipped average values of the the anisotropy in 1-$r_{\rm h}$-wide bins (steps of 0.25 $r_{\rm h}$; only bins with at least 5 points are considered). From top to bottom, the plots refer to clusters with Age/$t_{\rm rh} \ge 10$, $7 \le$ Age/$t_{\rm rh}<10$ and Age/$t_{\rm rh} < 7$. Left panels refer to non-core-collapsed clusters, while those on the right shows the results for core-collapsed clusters.}
\label{fig:anisotropy}
\end{figure*}

Figure~\ref{fig:trel} shows the ratio between the tangential and radial components of the velocity dispersions measured at the half-light radius as a function of the dynamical ages of the clusters as measured by the ratio of their physical age\footnote{Ages are mainly from \citet{2010ApJ...708..698D}. For NGC~1851, NGC~2808, NGC~6388, NGC~6441, NGC~6656 and NGC~6715, we considered the values from \citet{2014ApJ...785...21M}, while for NGC~5897 we refer to \citet{2014A&A...565A..23K}. Finally, the age of NGC~6791 is from \citet{2021A&A...649A.178B}.} to their half-mass relaxation time ($t_{r{\rm h}}$; from the Baumgardt catalog\footnote{For NGC~6791, we considered a half-light radius of 4.1 arcmin from \citet{2019MNRAS.483.2197K}.}). The values of $\sigma_{\rm tan}$ and $\sigma_{\rm rad}$ at the half-light radius were derived by fitting a 4-th order polynomial function to the corresponding profile as described in Sect.~\ref{kin}. We excluded from the plot all GCs for which the \textit{HST} data do not reach the half-light radius (so the anisotropy ratio would need to be extrapolated). Red points represent core-collapsed GCs, while all other systems are presented as black points. Core-collapsed GCs (regardless of whether they are considered as `possible' or `post' core-collapsed GCs) were labeled as such according to \citet{1995AJ....109..218T} or, if they were not included in that list, according to the Harris catalog\footnote{There are some discrepancies between the list of core-collapsed GCs in \citet{1995AJ....109..218T} and those in the Harris catalog, specifically NGC~6717 and NGC~6723. In the following, we consider the classification of \citet{1995AJ....109..218T}, for which NGC~6717 is a possible core-collapsed GC and NGC~6723 is not.}. Our analysis shows that dynamically older clusters (Age/$t_{\rm rh}$ $\gtrsim$ 10) tend to be isotropic even at the half-light radius, while dynamically-young systems are characterized by a radially-anisotropic velocity distribution at the half-light radius. A transition between the two regimes happens at Age/$t_{\rm rh}$ between 7 and 10. These findings are consistent with the theoretical expectations discussed above \citep[see, e.g., ][]{2014MNRAS.443L..79V,2016MNRAS.455.3693T,2017MNRAS.471.1181B} and previous observational works \citep{2015ApJ...803...29W}.\looseness=-4

To further illustrate the theoretical expectations concerning the evolution of the radial anisotropy, in Fig.~\ref{fig:trel} we show the time evolution of $\sigma_{\rm tan}/\sigma_{\rm rad}$ (calculated at the projected half-light radius) from a few Monte Carlo simulations run with the MOCCA code \citep{2013MNRAS.431.2184G}. The simulations follow the dynamical evolution of a few simple stellar systems composed of 500k stars with masses following a \citet{2001MNRAS.322..231K} initial mass function between 0.1 and 0.8 $M_{\odot}$, and spatially distributed according to the density profiles of \citet{1966AJ.....71...64K} models with values of the central dimensionless potential equal to $W_0=5$ and $W_0=7$ (corresponding, respectively, to $c \simeq 1.03$ and $c \simeq 1.53$). The systems are characterized by an initial anisotropic velocity distribution following the Osipkov-Merrit profile \citep[see, e.g.,][]{2008gady.book.....B}, $\beta = 1-\sigma_{\rm tan}^2/(2\sigma_{\rm rad}^2)=1/(1+r_{\rm a}^2/r^2)$ with $r_{\rm a}$ equal to the half-mass radius. As shown in this plot, the initial anisotropy of the clusters at the half-light radius gradually decreases during the cluster's long-term evolution. For the tidally filling system, the enhanced rate of star loss leads to a more rapid isotropization of the velocity dispersion. At the time when the system reaches core collapse, the cluster's radial anisotropy slightly increases, then continues its gradual decrease toward isotropy \citep[see also][for a study of the evolution of anisotropy for systems with a variety of different initial conditions]{2016MNRAS.455.3693T}. The differences between the anisotropy of systems that have similar dynamical ages but are in the pre- or post-core-collapsed phase is small, and within the uncertainty of the observed values.\looseness=-4

In Fig.~\ref{fig:anisotropy}, to further investigate the kinematic anisotropy in stellar clusters, we divided the sample into three groups, i.e., clusters with a Age/$t_{\rm rh} \ge 10$, between 7 and 10 (the Age/$t_{\rm rh}$ transition region found in Fig.~\ref{fig:trel}), and lower than 7. In each group, we also separated core-collapsed clusters (right panels) from all other systems (left panels). We collected all anisotropy measurements shown in Figs.~\ref{fig:kin1}--\ref{fig:kin4}, and plotted them as a function of distance from the center of the cluster in units of $r_{\rm h}$. Gray points are the individual measurements, while black dots are the moving averages of those points. Clusters with Age/$t_{\rm rh} \ge 10$ are isotropic at all distances within our FoV, regardless of their core-collapsed status, which is what we expected. Clusters with Age/$t_{\rm rh}$ between 7 and 10 are again shown to be isotropic at all distances, although the non-core-collapsed sample hints at a marginal radial anisotropy at $r \gtrsim r_{\rm h}$. Finally, dynamically-young clusters clearly present the expected radial anisotropy at large radii.\looseness=-4

Thus, clusters that underwent a core collapse seem to have similar velocity fields as those of the other GCs with similar dynamical ages; this appears to be consistent with what suggested by the results of the simulations presented in Fig.~\ref{fig:trel}, which show that core collapse has only a relatively small effect on the radial anisotropy measured at the half-light radius. Additional simulations and a larger observational sample of core-collapsed clusters are necessary to further explore this issue. In particular, it is worth noticing that the core-collapsed sample with Age/$t_{\rm rh} < 7$ is composed of only four clusters: NGC~6541, NGC~6752, NGC~7099 and NGC~7078. While the first three objects show an isotropic field even slightly farther than the half-light radius, the latter presents a strong radial anisotropy. This feature for NGC~7078 has also been noted by \citet{2014ApJ...797..115B} and \citet{2021MNRAS.505.5978V}. Among these four GCs, NGC~7078 (i) is more massive, (ii) is further from the center of the Galaxy, and (iii) has had fewer interactions with the Galactic potential of the Bulge/Bar and Disk \citep[see][]{2021MNRAS.505.5978V}. The combination of these properties could have preserved some of the original radial anisotropy in the innermost regions \citep[e.g.,][]{2014MNRAS.443L..79V}.\looseness=-4

\begin{figure}[t!]
\centering
\includegraphics[width=\columnwidth]{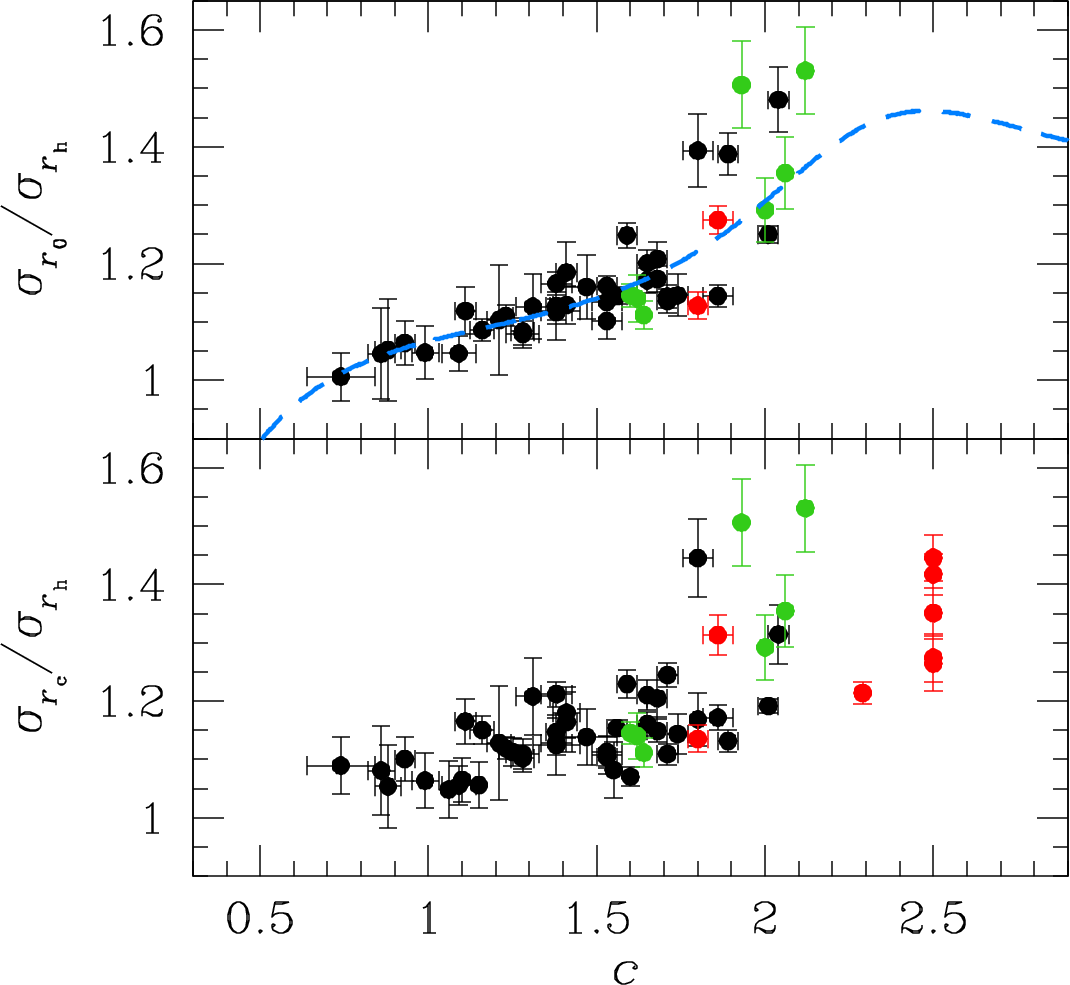}
\caption{(Bottom panel): ratio between the velocity dispersion at the core and half-light radii (from the Baumgardt catalog) as a function of concentration index $c$. The values of $c$ are obtained from \citet{2005ApJS..161..304M}. For GCs not analyzed in that work, we used the values from \citet[2010 edition]{1996AJ....112.1487H}. Red points are GCs that are marked as core-collapsed in the Harris catalog; green dots are measurements from \citet{2021AJ....161...41C}; black dots are all other objects. (Top panel): as in the bottom panel, but using the velocity dispersion at $r_0$ and $r_{\rm h}$ \citep[radii from][]{2005ApJS..161..304M}. Some core-collapsed GCs are missing from the plot because they were not analyzed by \citet{2005ApJS..161..304M}. The results obtained with the \texttt{\textsc{LIMEPY}} models are shown as an azure, dashed line.}
\label{fig:ratios}
\end{figure}

Core-collapsed GCs are not only more spatially concentrated than other GCs, but their velocity-dispersion radial profiles are also steeper. This has been shown by \citet{2015ApJ...803...29W} and \citet{2021AJ....161...41C}.  In Fig.~\ref{fig:ratios}, we provide an updated version of this finding, and a comparison with theoretical models. The bottom panel of Fig.~\ref{fig:ratios} shows the ratio between the velocity dispersion at the core and half-light radii ($\sigma_{r_{\rm c}}/\sigma_{r_{\rm h}}$), computed as described before at $r_{\rm c}$ and $r_{\rm h}$ from the Baumgardt catalog, as a function of the concentration index $c$. The values of $c$ are taken from \citet{2005ApJS..161..304M} and are defined as the $\log(r_{\rm t}/r_0)$, where $r_{\rm t}$ and $r_0$ are the tidal and King scale radii, respectively. We selected $c$ as obtained from the fit of a \citet{1966AJ.....71...64K} profile. If $c$ is not provided for a GC in the work of \citet{2005ApJS..161..304M}, we used the value in \citet[2010 edition]{1996AJ....112.1487H}. Red, filled circles mark core-collapsed GCs, while all other systems are plotted as black dots. Green points are clusters taken from the sample of \citet{2021AJ....161...41C}. This panel highlights a different location in the plot for core-collapsed GCs and the other stellar systems. This Figure confirms the trend between the spatial concentration and the steepness of the radial profile of the velocity dispersion.\looseness=-4

In order to further explore this trend and carry out a consistent comparison with theoretical models, we show in the top panel of Fig.~\ref{fig:ratios} the ratio of the values of the the velocity dispersions calculated at the King scale radius and at the half-light radius \citep[this time obtained at $r_0$ and $r_{\rm h}$ of][for consistency]{2005ApJS..161..304M} versus the concentration $c$ together with a line showing the expected variation of these quantities along the sequence of King models 
\citep[calculated using the \texttt{\textsc{LIMEPY}} software\footnote{\href{https://github.com/mgieles/limepy}{https://github.com/mgieles/limepy}};][]{2015MNRAS.454..576G}. The overall agreement between the observed and the theoretical sequences shown in the top panel of Fig.~\ref{fig:ratios} further supports the idea that more concentrated GCs have a steeper velocity-dispersion radial profiles according to what is expected for the family of King models with different values of $c$. It is interesting to notice the increasing deviation of the observational data from the theoretical King sequence at larger concentrations ($c \gtrsim 1.6$), including some of those from the sample of \citet[green points in Fig.~\ref{fig:ratios}]{2021AJ....161...41C}, which, according to those authors, are in an advanced dynamical state, being close to (or having recently undergone) core collapse. Although this issue requires further investigation, it may represent a kinematic manifestation of the deviation of clusters from the dynamical properties of King models as they approach core-collapse.\looseness=-4

There are a few exceptions in the overall picture described in the top panel of Fig.~\ref{fig:ratios}. The core-collapsed GC with $c<2.0$ and $\sigma_{r_{\rm c}}/\sigma_{r_{\rm h}}<1.2$ is NGC~362. NGC~362 is a post-core-collapsed GC \citep[e.g.,][]{2013ApJ...778..135D,2018ApJ...861...99L}, and its peculiar position could be explained by the structural and dynamical evolution of post-core-collapsed GCs being driven by gravothermal oscillations \citep[e.g.,][]{1996ApJ...471..796M}.\looseness=-4

Three clusters shown with black dots seem to have characteristics similar to core-collapsed GCs. These GCs are NGC~5272, NGC~6652 and NGC~6715. \citet{2005ApJS..161..304M} pointed out that NGC~5272 shows deviations from the classical King isotropic model, and can be better fit with other models \citep[see also][]{1976ApJ...206..128D,1979AJ.....84..752G}. NGC~6652 is not considered a core-collapsed GC, but it is very concentrated and has a very steep surface-brightness profile, typical of core-collapsed systems \citep{2006AJ....132..447N}. NGC~6715 is a cluster at the center of the Sagittarius Dwarf, an environment that could explain its peculiar core-collapsed-like concentration.\looseness=-4

\subsection{Kinematic distances}\label{kdist}

The Baumgardt's GC database also contains LOS velocity measurements. We made use of them to estimate distances of GCs by using the simple relation between the velocity dispersion along the LOS and in the plane of the sky:
\begin{equation}\label{eq:distance}
    \begin{aligned}
        \sigma_{\rm LOS} = 4.7404\,d\,\sigma_\mu\,;
    \end{aligned}
\end{equation}
where $d$ is the distance in kpc, $\sigma_{\rm LOS}$ and $\sigma_\mu$ are the velocity dispersions along the LOS (in km s$^{-1}$) and in the plane of the sky (from PMs in mas yr$^{-1}$), respectively, and 4.7404 (km\,yr\,kpc$^{-1}$\,mas$^{-1}$\,s$^{-1}$) is the conversion factor. The only assumption here is that the GCs are isotropic. We computed the distance $d$ by comparing the values of the velocity dispersion at the same distance, i.e., at the center of the cluster, obtained for both LOS and PM measurements by fitting a polynomial function to the corresponding velocity-dispersion radial profiles as described in Sect.~\ref{kin}. Our results are summarized in Table~\ref{tab:distance}. We considered only clusters with enough LOS velocities to solve the polynomial fit.\looseness=-4

\begin{figure}[t!]
\centering
\includegraphics[width=\columnwidth]{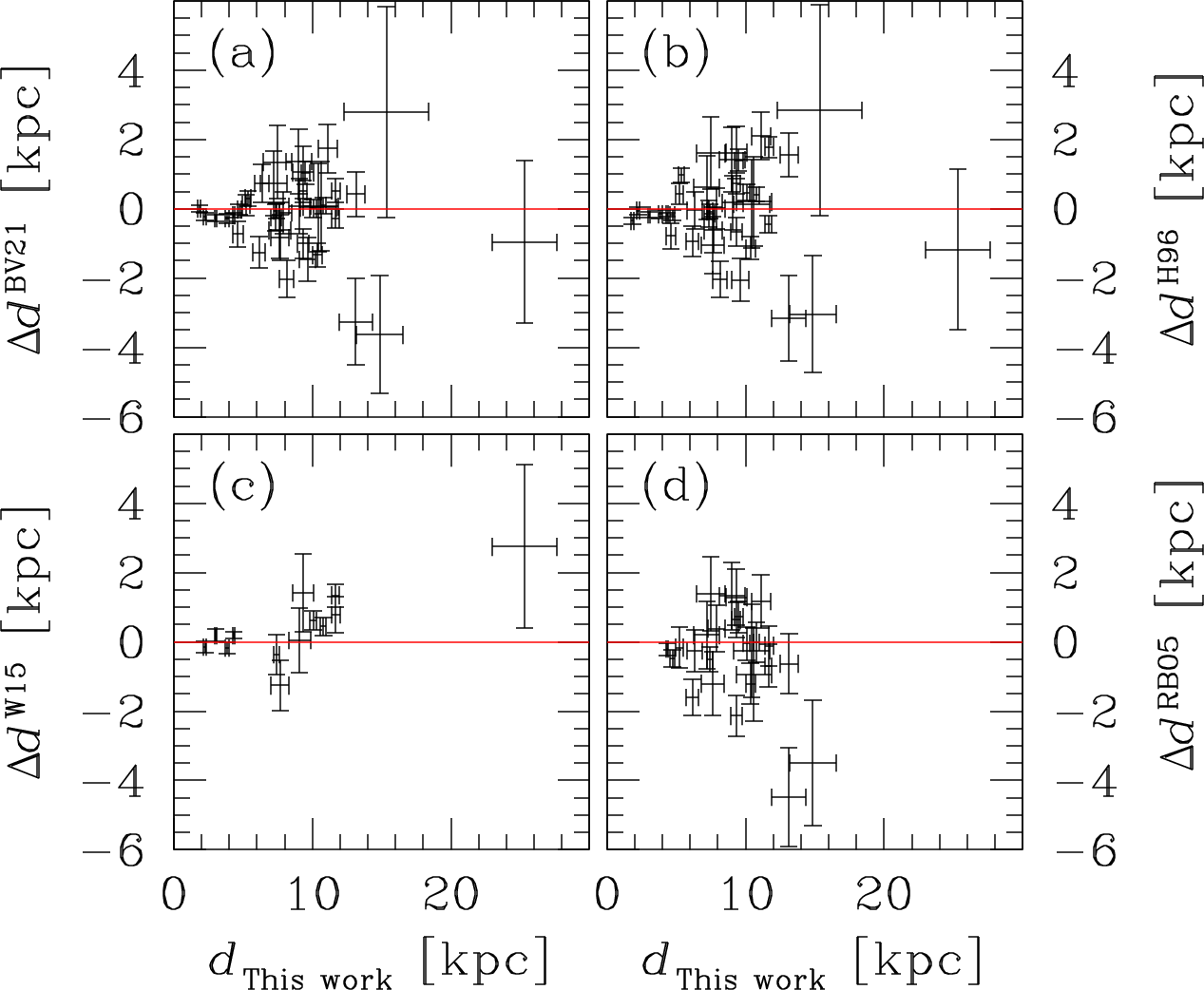}
\caption{Comparison between our distances and those in the literature: \citet[panel a]{2021MNRAS.505.5957B}, the Harris catalog (panel b), \citet[panel c]{2015ApJ...812..149W}  and \citet[panel d]{2005A&A...432..851R}. Only error bars are shown for clarity (no error bars are available for distances from the Harris catalog). The red lines are at 0.}
\label{fig:distance}
\end{figure}

Figure~\ref{fig:distance} shows comparisons between our distances and those from \citet[panel a; using \textit{Gaia} parallaxes, the comparison between PM and LOS velocities, and/or star counts]{2021MNRAS.505.5957B}, the Harris catalog (panel b; collected from various sources in the literature), the work of \citet[panel c; obtained with an approach similar to that used in our paper]{2015ApJ...812..149W} for a sample of 14 GCs, and the estimates from \citet[panel d; using the luminosity level of the zero-age horizontal branch\footnote{Distance moduli in \textit{HST} Wide-Field Planetary Camera 2 (WFPC2) F555W filter of \citet{2005A&A...432..851R} were converted to distances in kpc after correcting for extinction using the extinction coefficient $A_{\rm F555W}$ provided in \citet{1995PASP..107..156H} and the $E(B-V)$ reddening in the Harris catalog.}]{2005A&A...432..851R}. The median differences between $d$ in our work and those in the literature shown in Fig.~\ref{fig:distance} are $(-0.01 \pm 0.16)$ \citep{2021MNRAS.505.5957B}, $(-0.05 \pm 0.18)$ kpc (Harris catalog), $(-0.19 \pm 0.17)$ kpc \citep{2015ApJ...812..149W}, and $(-0.23 \pm 0.20)$ kpc \citep{2005A&A...432..851R}, respectively. All distance estimates are in agreement with these literature values at the $\sim$1$\sigma$ level. This is further proof of the goodness of our PM measurements. At large distances ($>$10 kpc), the differences between our distances and those from the literature increase, and so does the scatter of the points in Fig.~\ref{fig:distance}. The lower velocity dispersions of some of these clusters, as well as the larger uncertainties both in the PMs, LOS velocities and parallaxes, are likely the reasons for these discrepancies.\looseness=-4

\begin{table}[t!]
    \centering
    \caption{GC kinematic distances.}
    \label{tab:distance}
    \begin{tabular}{cc|cc}
    \hline
    \hline
    Cluster & $d$ & Cluster & $d$ \\
    & $[$kpc$]$ & & $[$kpc$]$ \\
    \hline
    NGC~104  & $ 4.34 \pm 0.06$ & NGC~6218 & $ 5.23 \pm 0.30$ \\
    NGC~288  & $ 9.08 \pm 0.79$ & NGC~6254 & $ 5.37 \pm 0.21$ \\
    NGC~362  & $ 9.33 \pm 0.31$ & NGC~6304 & $ 7.50 \pm 1.04$ \\
    NGC~1261 & $13.14 \pm 1.23$ & NGC~6341 & $ 7.68 \pm 0.66$ \\
    NGC~1851 & $11.66 \pm 0.25$ & NGC~6362 & $ 9.02 \pm 0.93$ \\
    NGC~2808 & $10.07 \pm 0.24$ & NGC~6388 & $11.69 \pm 0.30$ \\
    NGC~3201 & $ 4.73 \pm 0.15$ & NGC~6397 & $ 2.25 \pm 0.11$ \\
    NGC~4590 & $10.44 \pm 1.28$ & NGC~6441 & $13.15 \pm 0.63$ \\
    NGC~4833 & $ 7.21 \pm 0.93$ & NGC~6541 & $ 7.36 \pm 0.38$ \\
    NGC~5024 & $14.86 \pm 1.69$ & NGC~6624 & $ 7.91 \pm 0.59$ \\
    NGC~5272 & $ 8.16 \pm 0.52$ & NGC~6656 & $ 3.03 \pm 0.09$ \\
    NGC~5286 & $ 9.65 \pm 0.61$ & NGC~6681 & $11.10 \pm 0.68$ \\
    NGC~5897 & $15.35 \pm 3.04$ & NGC~6715 & $25.32 \pm 2.31$ \\
    NGC~5904 & $ 7.42 \pm 0.21$ & NGC~6723 & $ 7.65 \pm 0.81$ \\
    NGC~5927 & $ 9.33 \pm 0.75$ & NGC~6752 & $ 3.85 \pm 0.13$ \\
    NGC~5986 & $10.61 \pm 1.28$ & NGC~6809 & $ 4.62 \pm 0.37$ \\
    NGC~6093 & $ 9.34 \pm 0.41$ & NGC~7078 & $10.81 \pm 0.21$ \\
    NGC~6121 & $ 1.85 \pm 0.10$ & NGC~7089 & $10.36 \pm 0.33$ \\
    NGC~6171 & $ 6.36 \pm 0.54$ & NGC~7099 & $ 9.50 \pm 0.32$ \\   
    NGC~6205 & $ 6.16 \pm 0.44$ \\
    \hline
\end{tabular}
\end{table}

\section{Possible applications}

We choose the GC NGC~5904 to showcase some scientific applications enabled by our PMs\footnote{NGC 5904 shows one of the cleanest and best-defined rotation curves obtained for a GC so far \citep{2018ApJ...861...16L}, thus suggesting that no significant residual rotation affects its kinematics in the plane of the sky.}. Note that not all PM catalogs provide the same PM precision and overall quality, and some of the examples described below cannot be applied.\looseness=-4

\subsection{Internal kinematics of mPOPs}\label{mpops}

\begin{figure*}[t!]
\centering
\includegraphics[width=0.997\textwidth]{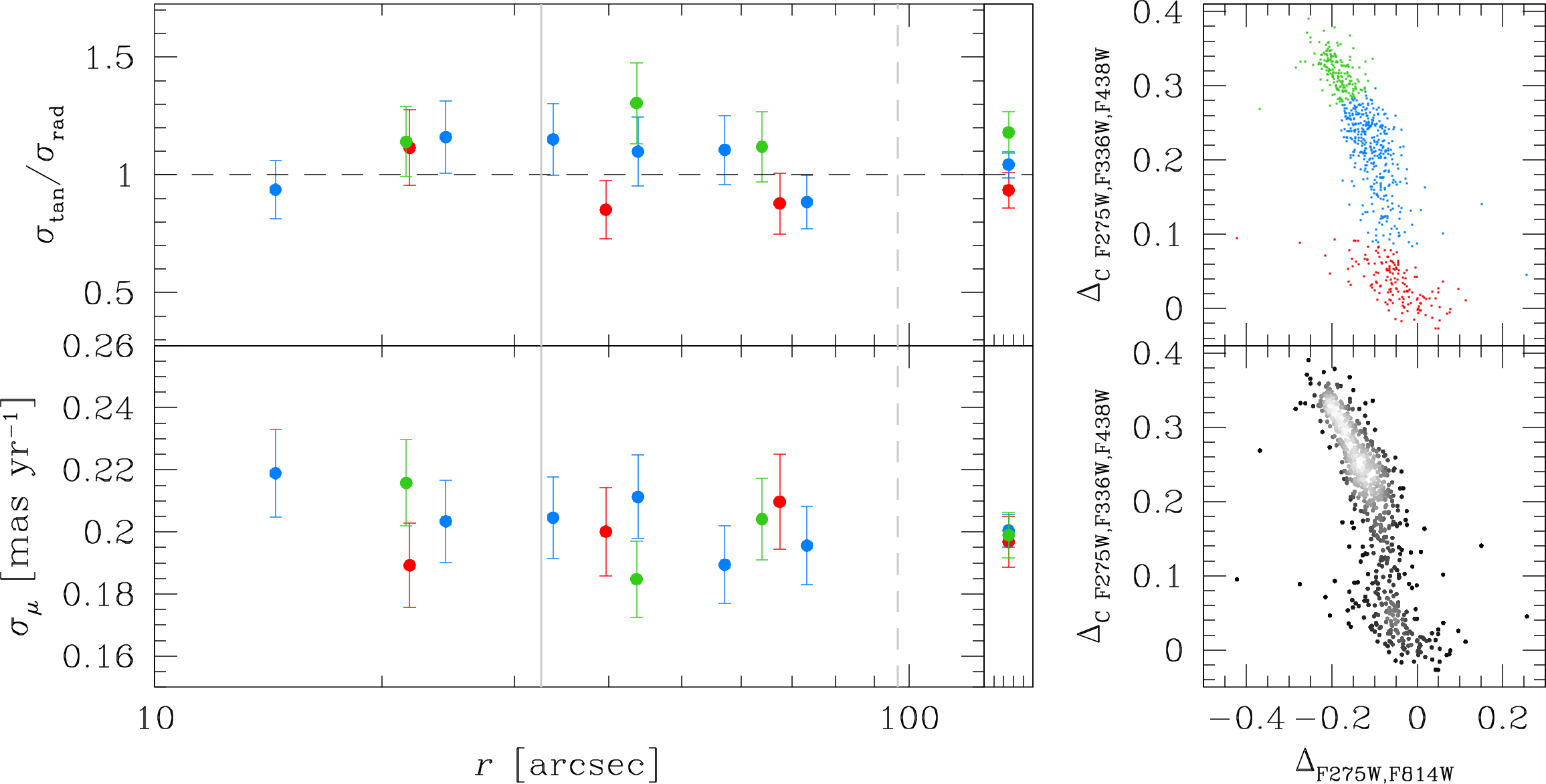}
\caption{Analysis of the internal kinematics of the mPOPs along the RGB of NGC~5904. We used the chromosome map of \citet{2017MNRAS.464.3636M} (top-right panel) and an Hess diagram (bottom-right panel) to select three groups of stars: 1G POPa (red points), 2G POPb and POPc (azure and green points, respectively). The left panels show the velocity-dispersion (bottom) and anisotropy (top) radial profiles for each population. 1G and 2G stars are kinematically isotropic and have the same velocity dispersions within our FoV. The gray solid and dashed vertical lines are set the core and half-light radii, respectively. The black, dashed horizontal line in the top-left panel is set to 1 (isotropic case). The small panels next to the left plots present the average velocity dispersion and anisotropy of each population across the entire FoV.\looseness=-4}
\label{fig:mpops}
\vspace{0.5 cm}
\includegraphics[width=0.997\textwidth]{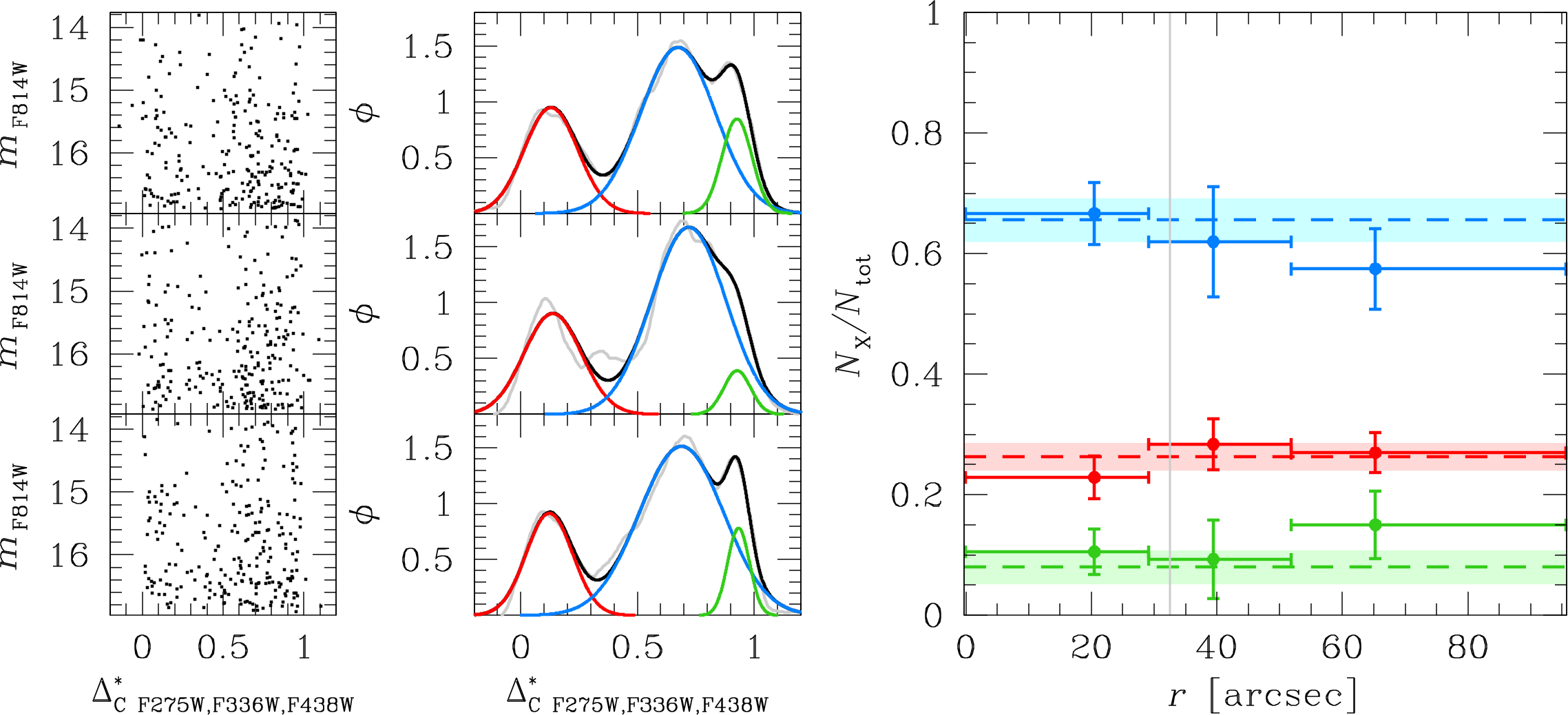}
\caption{Analysis of the spatial distributions of the mPOPs along the RGB of NGC~5904. The left panels show the $m_{\rm F814W}$ versus $\Delta^*_{\rm C\,F275W,F336W,F438W}$ CMDs for stars in each radial bin (with the distance from the center of the cluster increasing from bottom to top). The central panels present the corresponding kernel-density distributions (gray lines) and the triple-Gaussian functions (1G POPa: red; 2G POPb: azure; 2G POPc: green). The sum of the three Gaussians is plotted as a black line. The right panel shows the spatial distributions of the three mPOPs. The horizontal lines (and their shaded regions) correspond to the ratio of each mPOP using all the stars in the field. The vertical, gray line is set at $r_{\rm c}$ ($r_{\rm h}$ is outside the plot boundaries).\looseness=-4}
\label{fig:mpopspatdistr}
\end{figure*}

Kinematic differences between mPOPs can be a proxy of the different initial formation and evolution of first- (1G) and second-generation (2G) stars \citep{2010ApJ...724L..99B,2013ApJ...779...85M,2016ApJ...823...61M,2015MNRAS.450.1164H,2016MNRAS.455.3693T,2019MNRAS.487.5535T,2013MNRAS.429.1913V,2021MNRAS.502.4290V,2019MNRAS.489.3269C,2021MNRAS.502.1974S}. Recently, numerous observational efforts have investigated the kinematic properties of 1G and 2G stars to help us understand the mPOP phenomenon. We now know that 1G and 2G stars share similar kinematic features in dynamically-old GCs, or at least in the centermost regions of GCs where the two-body relaxation time is short, because two-body processes have already erased any kinematic differences between mPOPs \citep{2010ApJ...710.1032A,2018ApJ...861...99L,2019ApJ...873..109L}. The outermost regions of GCs are instead less relaxed, and some fingerprints of the initial kinematic differences between mPOPs can still be detected \citep{2013ApJ...771L..15R,2015ApJ...810L..13B,2018ApJ...853...86B,2018MNRAS.479.5005M,2018ApJ...859...15D,2018ApJ...864...33D,2019ApJ...884L..24D,2020ApJ...889...18C,2020ApJ...898..147C}.\looseness=-4

Evidence of the presence of mPOPs in NGC~5904 has been shown both spectroscopically \citep{2001AJ....122.1438I,2009A&A...505..117C,2013A&A...549A..41G} and photometrically \citep{2017MNRAS.464.3636M,2017ApJ...844...77L,2021ApJ...918L..24L}. To identify the mPOPs in our field, we made use of the pseudo color-color diagram (``chromosome map") computed by \citet{2017MNRAS.464.3636M} for this stellar cluster. We cross-correlated our PM catalog of NGC~5904 with that of \citet{2017MNRAS.464.3636M} and used their chromosome map to select 1G and 2G stars along the RGB of this cluster (right panels of Fig.~\ref{fig:mpops}). We identified three groups of stars: a 1G population (hereafter POPa in red) and two 2G groups (POPb and POPc in azure and green, respectively). The 1G-2G tagging was obtained in a similar way to what is shown in Fig.~4 of \citet{2017MNRAS.464.3636M}. The two 2G sub-populations were arbitrarily identified using the Hess diagram in Fig.~\ref{fig:mpops}.\looseness=-4

We then measured\footnote{In addition to the criteria described in Sect.~\ref{kin}, stars that were analyzed in this section passed the photometric criteria described in \citet{2017MNRAS.464.3636M}.} the velocity dispersions of each population in various radial bins of at least 50 stars each. The velocity-dispersion (bottom-left panel) and anisotropy (top-left panel) radial profiles show that 1G and 2G stars have similar kinematic temperature and are isotropic within our FoV (there is only a marginal hint of a radially and tangentially anisotropic POPa and POPc, respectively). This is expected given that our field covers out to about the half-light radius of the cluster, a region where two-body encounters have likely removed any initial kinematic differences between mPOPs. Our findings are in agreement with those obtained by \citet{2020ApJ...889...18C} with the \textit{Gaia} DR2 PMs.\looseness=-4

This result is in contrast with the finding of \citet{2021ApJ...918L..24L}, which measured the analog of the POPc stars as more spatially concentrated than the other two populations even within the half-light radius. However, analogously to the internal motions, differences in the spatial segregation of mPOPs can be preserved in regions where the relaxation time is long enough to preserve them. Thus, the complete mixing and similar kinematic features are likely expected for the mPOPs in the core of NGC~5904.\looseness=-4

To shed light on this disagreement, we computed the spatial distribution of the three populations in our FoV as follows. First, we divided our sample\footnote{We removed the constraint on the PM error to increase the statistics at our disposal.}  of RGB stars into three equally-populated bins and computed a kernel-density distribution of the $\Delta^*_{\rm C\,F275W,F336W,F438W}$ color for the stars in each bin. The $\Delta_{\rm C\,F275W,F336W,F438W}$ color was obtained \citet{2017MNRAS.464.3636M} after rectifying the RGB sequences in the $m_{\rm F814W}$ versus ${\rm C_{F275W,F336W,F438W}}$ CMD. However, the sequences have been rectified using all stars in the FoV and, in different radial bins, they can show some deviations from being exactly vertical. Thus, we fine-tuned the $\Delta_{\rm C\,F275W,F336W,F438W}$ color in each radial bin by using two fiducial lines (one each on the red and blue sides of the RGB sequence, respectively) drawn by hand and then computing a new color $\Delta^*_{\rm C\,F275W,F336W,F438W}$ as in \citet[see left panels of Fig.~\ref{fig:mpopspatdistr}]{2019ApJ...873..109L}:\looseness=-4
\begin{equation}
    \begin{split}
        \Delta^*_{\rm C\,F275W,F336W,F438W} = \\
        = \frac{(\Delta_{\rm C\,F275W,F336W,F438W} - \rm fiducial_{red})}{(\rm fiducial_{blue} - \rm fiducial_{red})} \, .
    \end{split}
\end{equation}

The kernel-density distribution\footnote{The kernel-density estimation was obtained with the \texttt{python} dedicated tools in \texttt{scikit-learn} \citep{scikit-learn}, and by assuming an `Epanechnikov' kernel with a bandwidth of 0.1. These parameters were chosen as a good compromise between smoothness and the preservation of the features in the mPOP distributions.} of the $\Delta^*_{\rm C\,F275W,F336W,F438W}$ color in each bin was fitted with a triple-Gaussian function (central panels of Fig.~\ref{fig:mpopspatdistr}), and the fraction of stars in each population was estimated following the same approach of \citet{2013ApJ...765...32B}. To ensure the density distribution was robust against the small statistics along the RGB, we obtained the fraction of each mPOP by bootstrapping with replacements the sample of stars 1\,000 times. The final values for the fractions of stars and their errors were determined as the median and the 68.27-th percentile of the distribution about the median, respectively. The ratios of the three mPOPs are shown in the right panel of Fig.~\ref{fig:mpopspatdistr}. At odds with the finding of \citet{2021ApJ...918L..24L}, we can see that the fractions of each mPOP do not vary as a function of distance from the center of the cluster within our FoV, as expected.\looseness=-4

\subsection{Energy equipartition}\label{eeq}

The energy-equipartition state of a GC is one of the most challenging measurements to obtain because it requires precise PMs along a (relatively) wide range of masses, i.e., for faint stars. Nowadays, this is one of the few applications that only \textit{HST} can allow us to investigate in detail.\looseness=-4

In collisional systems in a certain state of energy equipartition, there is a relation between the stellar mass $m$ and the velocity dispersion $\sigma_\mu$: $\sigma_\mu \propto m^{-\eta}$, where $\eta$ is the level of energy equipartition of the system. Theoretical works \citep{2013MNRAS.435.3272T,2016MNRAS.458.3644B,2017MNRAS.464.1977W} have shown that a complete state of energy equipartition ($\eta = 0.5$) is never reached because of the so-called Spitzer instability. Recent observational works on this topic have verified the goodness of this prediction \citep{2010ApJ...710.1032A,2018ApJ...853...86B,2018ApJ...861...99L,2019ApJ...873..109L}.\looseness=-4

We made use of the exquisite PMs of NGC~5904 to measure its state of energy equipartition using both the parameter $\eta$ defined above and the equipartition mass parameter $m_{\rm eq}$ introduced by \citet{2016MNRAS.458.3644B}\footnote{According to \citet{2016MNRAS.458.3644B}, the relation between the velocity dispersion $\sigma_\mu$ and the stellar mass is as follows:
\begin{equation*}
  \sigma (m) = \left\{
  \begin{array}{lc}
      \sigma_0 \exp(-\frac{1}{2}\frac{m}{m_{\rm eq}}) & {\rm if}~{m \le m_{\rm eq}} \\
      \sigma_0 \exp(-\frac{1}{2}) (\frac{m}{m_{\rm eq}})^{-\frac{1}{2}} & {\rm if}~{m > m_{\rm eq}} \\
  \end{array}
  \right. \, ,
\end{equation*}
where $\sigma_0$ is the velocity dispersion at $m = 0$. To avoid unphysical values of the level of energy equipartition, there is cutoff at $m = m_{\rm eq}$.}. As a reference, a high degree of energy equipartition is characterized by a large value of $\eta$ and a small value of $m_{\rm eq}$.\looseness=-4

\begin{figure*}[t!]
\centering
\includegraphics[width=0.99\textwidth]{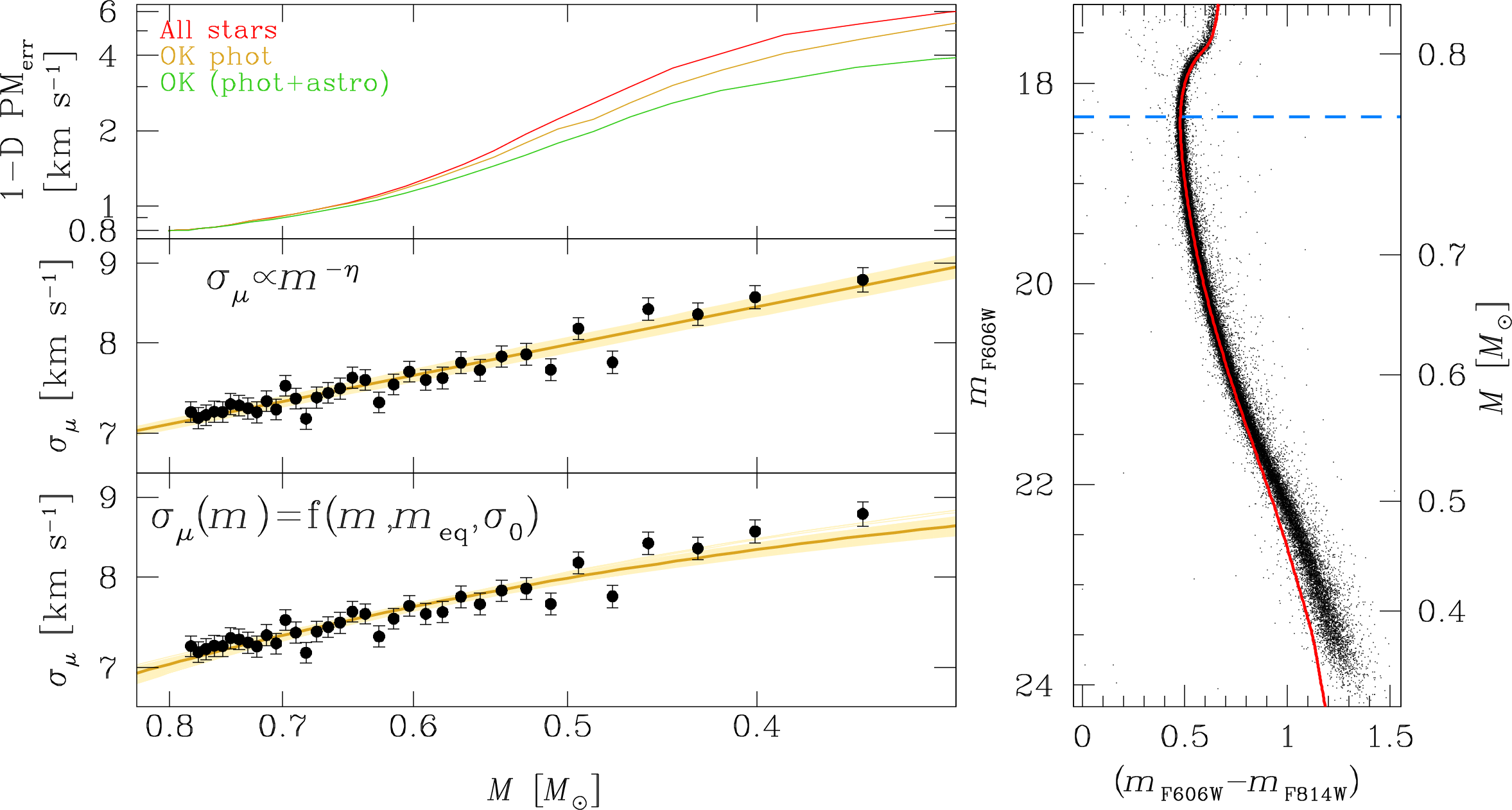}
\caption{Measurement of the level of energy equipartition in NGC~5904. In the CMD on the right, we show in red the isochrone used to estimate the mass of the stars along the MS. In the left panels, we show the velocity dispersion $\sigma_\mu$ as a function of mass for MS stars. The gold line is the best fit obtained with the scale parameter $m_{\rm eq}$ (bottom panel) or the classical formalism of $\eta$ (middle panel). The yellow regions correspond to the 1$\sigma$ errors of the fit. The top-left panel presents the median 1-D PM error as a function of stellar mass for three samples: all objects in the catalog (red line), sources that passed our photometric quality selections (yellow line), and stars that survived both the photometric and astrometric cuts (green line).\looseness=-4}
\label{fig:eeq}
\includegraphics[width=0.99\textwidth]{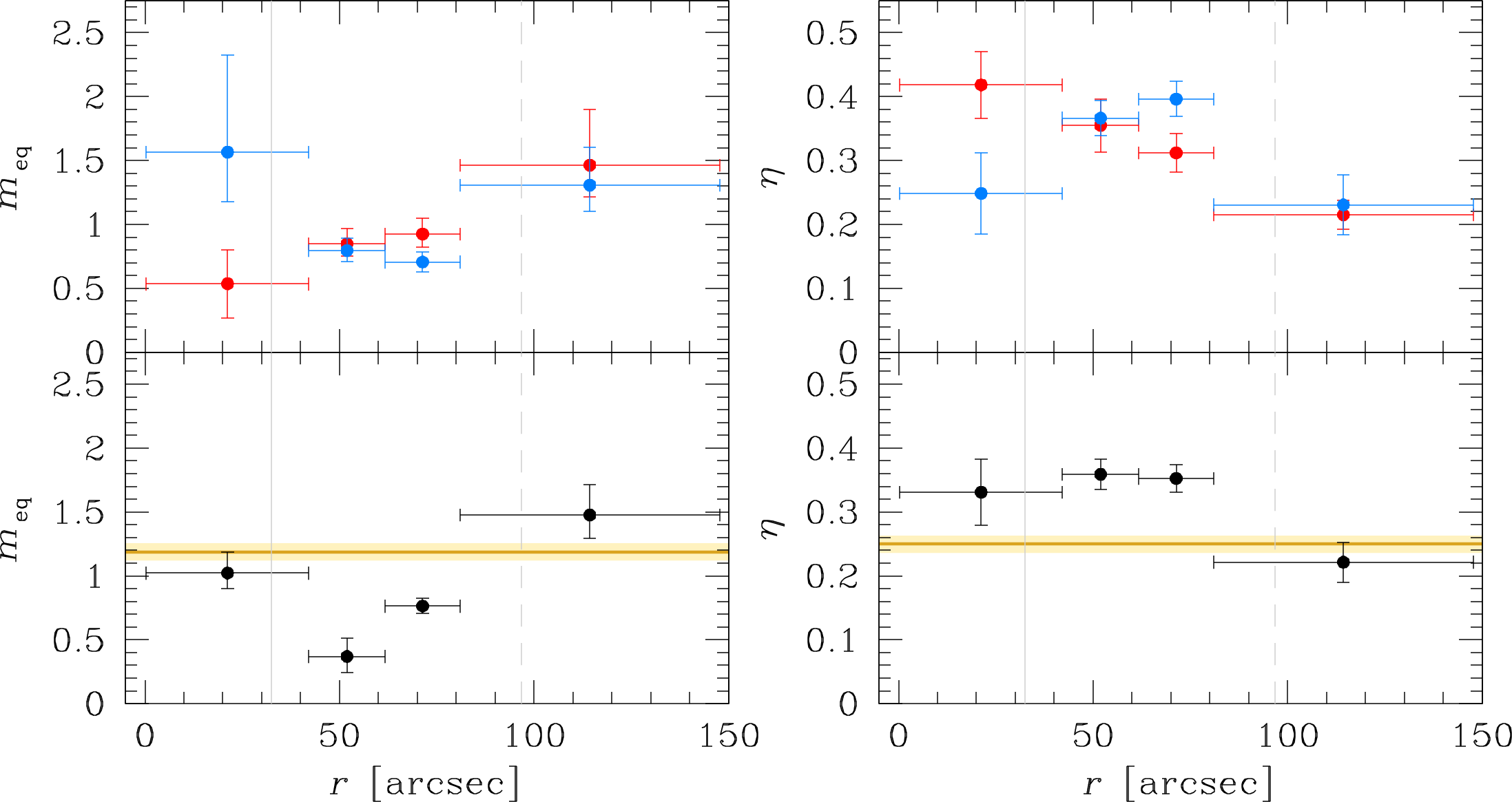}
\caption{Local levels of energy equipartition of NGC~5904. Left panels show the value of $m_{\rm eq}$ as a function of distance from the center of the cluster, while right panels present the same trend for $\eta$. Bottom panels are obtained by considering the combined velocity dispersion $\sigma_\mu$. The gold line and the yellow region represent the global value of the level of energy equipartition and its 1$\sigma$ region as measured in Fig.~\ref{fig:eeq}. In the top panels, we highlight the trends obtained by estimating the level of energy equipartition using either the radial component $\sigma_{\rm rad}$ (red points) or the tangential component $\sigma_{\rm tan}$ (blue points).\looseness=-4}
\label{fig:eeq_rad}
\end{figure*}

We started by inferring the mass of stars along the MS below the MS turnoff by means of the updated isochrones of the ``A Bag of Stellar Tracks and Isochrones" \citep[BaSTI;][]{2018ApJ...856..125H}. The parameters for NGC~5904 were chosen from the recent work of \citet{2019MNRAS.483.4949G}: a solar-scaled, 12.15 Gyr old isochrone for $[$Fe$/$H$] = -1.33$, $Y = 0.2478$, accounting for atomic diffusion, with a distance of 7.4 kpc. The fit is shown in Fig.~\ref{fig:eeq}. Although not perfect for the faintest portion of the MS due to the still-existing shortcomings in this mass regime of the color-effective temperature relationship, the fit is good enough to assign a mass to each star.\looseness=-4

We then computed the $\sigma_\mu$ in 35 bins of 1077 stars each along the MS. The bottom-left panel of Fig.~\ref{fig:eeq} shows the result obtained by fitting the mass-dependent exponential relation of \citet{2016MNRAS.458.3644B} with  a  maximum-likelihood approach. We find a global $m_{\rm eq} = 1.18 \pm 0.07$ $M_\odot$. \citet{2016MNRAS.458.3644B} provided a relation between $m_{\rm eq}$ and the ratio between the cluster age and its core relaxation time. Using the age of \citet{2019MNRAS.483.4949G} and a core relaxation time of 0.19 Gyr \citep[2010 edition]{1996AJ....112.1487H}, we expect a value of $m_{\rm eq} = 1.67^{+0.32}_{-0.28}$ $M_\odot$, which is in agreement with our estimate at the 2$\sigma$ level.\looseness=-4

The middle-left panel presents the linear fit in the same log-log plane. We found a value of $\eta = 0.25 \pm 0.01$. As for the value of $m_{\rm eq}$, this finding is consistent with the theoretical predictions of a partial state of energy equipartition even in an advanced stage of the cluster evolution.\looseness=-4

As a reference, the median 1-D PM error as a function of stellar mass is plotted in the top-left panel of Fig.~\ref{fig:eeq}. The three lines correspond to the median PM trends of: all objects in the catalog (red line), sources that passed our photometric quality selections (see Sect.~\ref{kin}; yellow line), and stars that survived both the photometric and astrometric cuts (always described in Sect.~\ref{kin}; green line).

Our data set allowed us to push this investigation even further. We divided our FoV in four equally-populated radial bins, and measured the level of energy equipartition in each bin using the velocity dispersions measured in ten equally-populated magnitude bins. The result is shown in the bottom panels of Fig.~\ref{fig:eeq_rad} (left for $m_{\rm eq}$ and right for $\eta$). We can notice marginal evidence of the innermost regions of the cluster having a higher degree of energy equipartition (low $m_{\rm eq}$ and high $\eta$) than the outskirts. The innermost point within $r_{\rm c}$ is the most uncertain because the fit is obtained with a smaller mass baseline.\looseness=-4

In the top panels, we reproduce a similar analysis, but this time measuring the level of energy equipartition using the radial and tangential components of the velocity dispersion separately. We find that the levels of energy equipartition from $\sigma_{\rm rad}$ and $\sigma_{\rm tan}$ are consistent with each other at all radii, with only marginal differences at the 1--2$\sigma$ level. The larger difference between the level of energy equipartition in the two components is shown in the innermost bin. The discrepancy is mainly the result of a poor fit, especially in the case of the exponential fit. This is in agreement with the simulations of \citet{2021MNRAS.504L..12P,2022MNRAS.509.3815P} who found the degree of energy equipartition in the two velocity components to be similar in the region within the half-light radius. Future studies will extend the investigation of the energy equipartition in the tangential and radial dispersions to the outer regions of GCs where, according to \citet{2021MNRAS.504L..12P,2022MNRAS.509.3815P}, the degrees of energy equipartition in these two velocity components may differ.\looseness=-4

\begin{figure*}[t!]
\centering
\includegraphics[width=\textwidth]{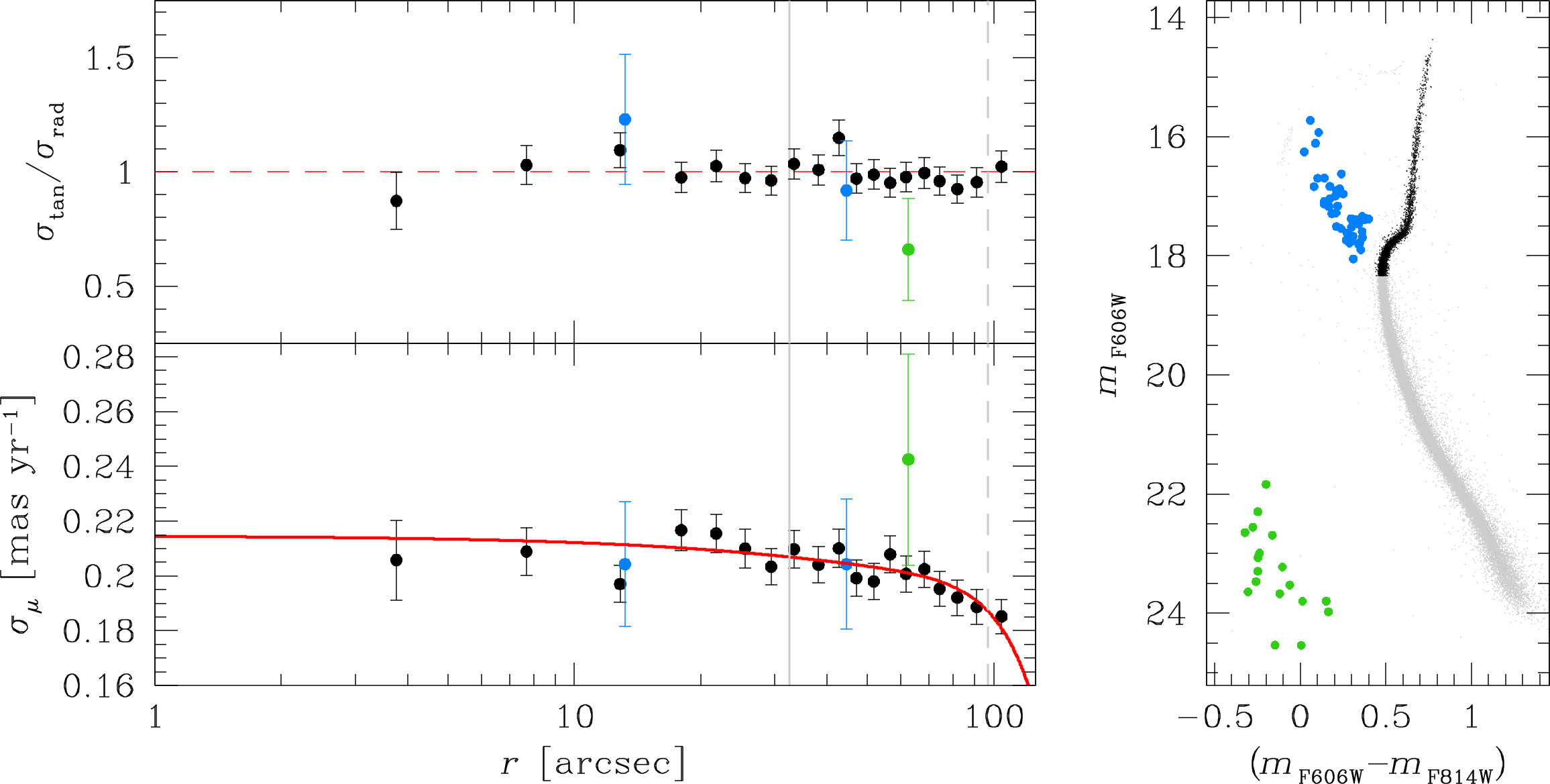}
\caption{The optical CMD of NGC~5904 is shown in the right panel. Black dots are the reference population used in the analysis (Sect.~\ref{weight}), blue dots are BSs, and green points correspond to WDs. All other objects are shown as gray dots. In the bottom-left panel, we show the velocity-dispersion radial profiles of the the reference population (in black), BSs (in blue) and WDs (in green). The red line is the polynomial fit to the $\sigma_\mu$ of the reference population. The top-left panel presents the anisotropy radial profile for the same stars. The red, dashed horizontal line is set to 1. In all the left panels of this Figure, the vertical lines mark the $r_{\rm c}$ (solid) and $r_{\rm h}$ (dashed).\looseness=-4}
\label{fig:bss_wd}
\end{figure*}

\paragraph{A note on the PM errors for faint stars} Despite our careful data reduction and PM computation, the PM errors might be under/overestimated. For example, a higher velocity dispersion for faint stars could be caused by underestimated PM uncertainties. We repeated the analysis shown in Fig.~\ref{fig:eeq} by using only stars brighter than (1) $m_{\rm F606W} = 22.1$ ($\sim 0.5 M_\odot$) and (2) $m_{\rm F606W} = 23.2$ ($\sim 0.42 M_\odot$). We find:\looseness=-4
\begin{itemize}
    \item $m_{\rm F606W} \le 22.1$ ($M \gtrsim 0.5 M_\odot$): $m_{\rm eq} = 1.60 \pm 0.18$ $M_\odot$ , $\eta = 0.21 \pm 0.02$ ;
    \item $m_{\rm F606W} \le 23.2$ ($M \gtrsim 0.42 M_\odot$): $m_{\rm eq} = 1.31 \pm 0.10$ $M_\odot$ , $\eta = 0.24 \pm 0.02$ .
\end{itemize}
Although the results are in agreement at the 1--2$\sigma$ level, it seems that the value of $m_{\rm eq}$ is larger ($\eta$ is smaller) the brighter is the magnitude cut, i.e., when we exclude the faintest objects with large PM errors. However, the exponential fit is less robust when the mass interval is small (the associated errors increase). The variation of $\eta$ is instead mild.\looseness=-4

Furthermore, the selections applied can bias the analyzed sample and the inferred kinematics, as extensively discussed in \citet{2014ApJ...797..115B}. In the case of NGC~5904, we find that the selection on the PM reduced $\chi^2$ applies a more significant cut on the PM errors, as shown in the top panel of Fig.~\ref{fig:eeq} by the comparison of the yellow and green lines. As a test, we repeated the same analysis as before upon rescaling the PM errors of the stars used in the measurement of the level of energy equipartition by the ratio between the median PM errors obtained with sources that passed our photometric quality selections (yellow line in the top panel of Fig.~\ref{fig:eeq}), and with stars that survived both the photometric and astrometric cuts (green line). We find a global level of energy equipartition of $m_{\rm eq} = 1.50 \pm 0.11$ $M_\odot$ and $\eta = 0.20 \pm 0.01$. Since we increased the PM errors, the intrinsic velocity dispersion of faint stars decreased, but again all estimates are in agreement within 2--3$\sigma$.\looseness=-4

It is hard to say if these tests are a proxy of underestimated PM errors for faint stars,if they are biased by the less-constrained fit, or if what we see is just due to the intrinsic kinematics of our tracers. Nevertheless, these examples highlight how important it is to understand the data set used. Once again, we advise users to carefully check the PMs, and test their quality selections on a cluster-by-cluster basis, especially when the PM errors are of the order of the intrinsic velocity dispersion of the stars.\looseness=-4

\subsection{Kinematic mass determination}\label{weight}

Knowledge of the relation between mass and velocity dispersion allows us to measure the kinematic mass of stars, similar to what has been done by \citet{2016ApJ...827...12B} and \citet{2018ApJ...861...99L,2019ApJ...873..109L}. The unknown mass of a group of stars X ($M_{\rm X}$) with velocity dispersion $\sigma_{\rm X}$ can be obtained as follows:
\begin{equation}
    \alpha_{\rm X} = \frac{\sigma_{\rm X}}{\sigma_{\rm ref}} = \left(\frac{M_{\rm X}}{M_{\rm ref}}\right)^{-\eta} \, ,
\end{equation}
where $M_{\rm ref}$ and $\sigma_{\rm ref}$ are the known mass and velocity dispersion of a reference population. For this analysis, we choose to measure the mass of two groups of stars: blue stragglers (BSs) and white dwarfs (WDs). Our reference population is the stars brighter than MS turnoff along the SGB and RGB (the same stars analyzed in Fig.~\ref{fig:kin2}), for which we assume $M_{\rm ref}$ to be equal to the mass of stars at the MS turnoff. Using the isochrone in Sect.~\ref{eeq}, we define $M_{\rm ref} = 0.78$ $M_\odot$.\looseness=-4

We selected BSs and WDs in various optical and UV CMDs (right panel of Fig.~\ref{fig:bss_wd}). We split the BSs into two groups (of 22 and 21 stars, respectively) and WDs into one group (of 14 stars), and measured their velocity dispersions $\sigma_\mu$. The result is shown in the bottom-left panel of Fig.~\ref{fig:bss_wd}. Blue points depict the velocity dispersions of the BSs, while the green dot refers to the WD kinematics. Black points are the reference population. The effect of the energy equipartition is shown in the plot: BSs, which are more massive than RGB, SGB and MS stars, seem to be slightly kinematically colder (lower $\sigma_\mu$) than the reference population. WDs have a mass of about 0.5 $M_\odot$ \citep[e.g.,][and references therein]{2019MNRAS.488.3857B}. Being less massive than the reference population, their $\sigma_\mu$ is instead marginally higher. Finally, the top-left panel shows that BSs and WDs are kinematically isotropic as the other stars at the $\sim$1$\sigma$ level.\looseness=-4

We previously fit the $\sigma_\mu$ of the reference population with a polynomial function (see Sect.~\ref{kin}). To obtain the mass of the BSs, we repeated the same computation but we also solved for $\alpha_{\rm BS}$ by rescaling the polynomial of the reference population to fit the BS $\sigma_\mu$ at the same time. By assuming the global value of $\eta = 0.25 \pm 0.01$, we obtain $M_{\rm BS} = 0.84 \pm 0.28$ $M_\odot$. However, we notice that the BSs are preferentially located in regions where the local level of energy equipartition is higher than the global level. If we use the average value of $\eta$ in these regions ($0.34 \pm 0.06$), the mass becomes $M_{\rm BS} = 0.82 \pm 0.20$ $M_\odot$. The difference between the two values is small since the value of $\alpha_{\rm BS}$ is close to unity. Both estimates are in agreement with the BS mass of NGC~5904 obtained by \citet[$0.82^{+0.29}_{-0.18}$ $M_\odot$]{2016ApJ...827...12B}. The mass of the BSs in this cluster is slightly lower than the average BS mass \citep[1.0--1.6 $M_\odot$; e.g.,][]{2018ApJ...860...36F}. However, it is worth noticing that our sample is mainly composed of BSs close to the MS turnoff, which are less massive than those on the bright-end of the BS sequence \citep[e.g.,][]{2019ApJ...879...56R}.\looseness=-4

For the WDs, a robust analysis cannot be obtained because we have only one point at our disposal for deriving $\alpha_{\rm WD}$. Nevertheless, we repeated the same analysis to qualitatively assess the mass of the WDs. If we use the global value of $\eta$, we obtain $M_{\rm WD} = 0.39 \pm 0.25$ $M_\odot$, which is lower than the average mass of the WDs in GCs. However, if we assume the local level of energy equipartition at the average WD distance from the center of the cluster, we find $M_{\rm WD} = 0.48 \pm 0.21$ $M_\odot$.\looseness=-4

\section{Conclusions}

We computed PMs for 57 stellar clusters studied in the GO-13297 program. The astro-photometric catalogs taht we have made publicly available represent the most complete, homogeneous collection of PMs of stars in the cores of stellar clusters to date, and more than double the number of clusters for which high-precision, \textit{HST}-based PMs are available \citep{2014ApJ...797..115B}. Furthermore, the astrometric information that we are releasing is complementary to that provided by the current (and future) \textit{Gaia} data releases. At the dawn of a new era in astronomy with the first light of the \textit{James Webb Space Telescope} (\textit{JWST}), the legacy that these PM catalogs offer is further enhanced, since they can serve as an important astrometric benchmark for \textit{JWST}-based data reduction and tools.\looseness=-4

We described the data reduction and, in great detail, the quality selections needed to select reliable objects for any kinematic analysis. We stress again that the data used for each cluster are different, thus any correction or selection should be tailored on a cluster-by-cluster basis. This is particularly important for stars with PM errors of the same order as the amplitude of the kinematic features that one wants to measure, for example, for stars along the MS of GCs.\looseness=-4

We made use of our catalogs to study the general kinematic properties of the bright, massive stars in our clusters. We provided additional evidence supporting early findings that dynamically-young systems have a radially anisotropic velocity distribution at the half-light radius, while in dynamically older clusters the velocity distribution is isotropic at the same distance from the center of the cluster. This trend is consistent with the theoretical results of simulations showing that initially radially anisotropic clusters evolve toward an isotropic velocity distribution during their long-term evolution. Interestingly, core-collapsed clusters show similar properties to the non-core-collapsed systems, although a larger sample of core-collapsed GCs will be necessary to confirm the similarities with non-core-collapsed clusters (in particular for the group with longer relaxation times).\looseness=-4

Finally, we showcased our PM catalogs using the GC NGC~5904. We separated the mPOPs along the RGB of the cluster and showed that, within our FoV, 1G and 2G stars have the same kinematics, are kinematically isotropic and have the same flat radial distributions. A detailed analysis of the kinematics of mPOPs will be the subject of another paper in this series.\looseness=-4

We investigated in detail the level of energy equipartition of NGC~5904. This cluster is in an advanced stage of its dynamical evolution, yet it has reached only a partial state of energy equipartition, as predicted by theoretical simulations. Knowledge of the level of energy equipartition also allowed us to measure the kinematic masses the BSs and WDs, finding a good agreement with the typical masses of these objects obtained with different methods in the literature.\looseness=-4

\clearpage

\section*{Acknowledgments}

The authors thank the anonymous referee for the detailed suggestions that improved the quality of our work. ML and AB acknowledges support from GO-13297, GO-15857 and GO-16298. EV acknowledges support from NSF grant AST-2009193. AA acknowledges support from the Spanish Agencia Estatal del Ministerio Ciencia e Innovaci\'on (AEI-MICINN) under grant PID2020-115981GB-I00. LB acknowledges the funding support from Italian Space Agency (ASI) regulated by ``Accordo ASI-INAF n. 2013-016-R.0 del 9 luglio 2013 e integrazione del 9 luglio 2015 CHEOPS Fasi A/B/C". LRB acknowledges support by MIUR under PRIN program \#2017Z2HSMF and PRIN-INAF-2019-PI:BEDIN. SC acknowledges support from INFN (Iniziativa specifica TAsP), and from PLATO ASI-INAF agreement n.2015-019-R.1-2018. FRF, ED e BL acknowledge funding from Italian MIUR throughout the PRIN-2017 grant awarded to the project Light-on-Dark (PI:Ferraro) through contract PRIN-2017K7REXT. ML thanks Drs. Peter Zeidler, Laura Watkins and Silvia Raso for the useful discussions on various aspects of this project.

This research was pursued in collaboration with the HSTPROMO (High-resolution Space Telescope PROper MOtion) collaboration, a set of projects aimed at improving our dynamical understanding of stars, clusters and galaxies in the nearby Universe through measurement and interpretation of proper motions from \textit{HST}, \textit{Gaia}, and other space observatories. We thank the collaboration members for the sharing of their ideas and software.

Based on observations with the NASA/ESA \textit{HST}, obtained at the Space Telescope Science Institute, which is operated by AURA, Inc., under NASA contract NAS 5-26555. This work has made use of data from the European Space Agency (ESA) mission {\it Gaia} (\url{https://www.cosmos.esa.int/gaia}), processed by the {\it Gaia} Data Processing and Analysis Consortium (DPAC, \url{https://www.cosmos.esa.int/web/gaia/dpac/consortium}). Funding for the DPAC has been provided by national institutions, in particular the institutions participating in the {\it Gaia} Multilateral Agreement. This research made use of \texttt{astropy}, a community-developed core \texttt{python} package for Astronomy \citep{astropy:2013, astropy:2018}, \texttt{\textsc{LIMEPY}} \citep{2015MNRAS.454..576G}, \texttt{emcee} \citep{2013PASP..125..306F}, \texttt{scikit-learn} \citep{scikit-learn}, and of the SIMBAD database \citep{2000A&AS..143....9W}, operated at CDS, Strasbourg, France.

\appendix

\setcounter{table}{0}
\renewcommand{\thetable}{A\arabic{table}}

\setcounter{figure}{0}
\renewcommand{\thefigure}{A\arabic{figure}}

\section{Absolute PMs}\label{abspm}

We cross-correlated each of our PM catalogs with the \textit{Gaia} EDR3 catalog and computed the PM zero-points to transform our relative PMs onto an absolute system.\looseness=-4

We considered only cluster members with well-measured \textit{HST} PMs (see Sect.~\ref{kin}) and whose \textit{Gaia} EDR3 PMs have an \texttt{RUWE} better than 1.25, an astrometric excess noise less than 0.4, a number of bad along-scan observations less than 1.5\% of the total number of along-scan observations, a PM error in each coordinate better than 0.1 mas yr$^{-1}$, and $G>13$. If fewer than 25 objects were found in common with our \textit{HST} sample, we relaxed these parameters to increase the statistics. The PM zero-point in each coordinate is defined as the 3.5$\sigma$-clipped average value of the difference between the \textit{HST} and \textit{Gaia} PMs. We set the error equal to the error of the mean. We also added in quadrature to our uncertainties a systematic error for the \textit{Gaia} EDR3 PMs of 0.026 mas yr$^{-1}$ \citep[obtained using Eq.~2 of][assuming an angular separation $\theta = 0$ deg since we are analyzing the clusters' cores]{2021MNRAS.505.5978V}. We do not take into account the rotation of the clusters in the plane of the sky, which is included in the PMs from the \textit{Gaia} catalog but not in those made with the \textit{HST} data (Sect.~\ref{pms}), although the scatter due to this effect is small. All values are reported in Table~\ref{tab:abspm}.\looseness=-4

\begin{figure*}
\centering
\includegraphics[width=0.9\textwidth]{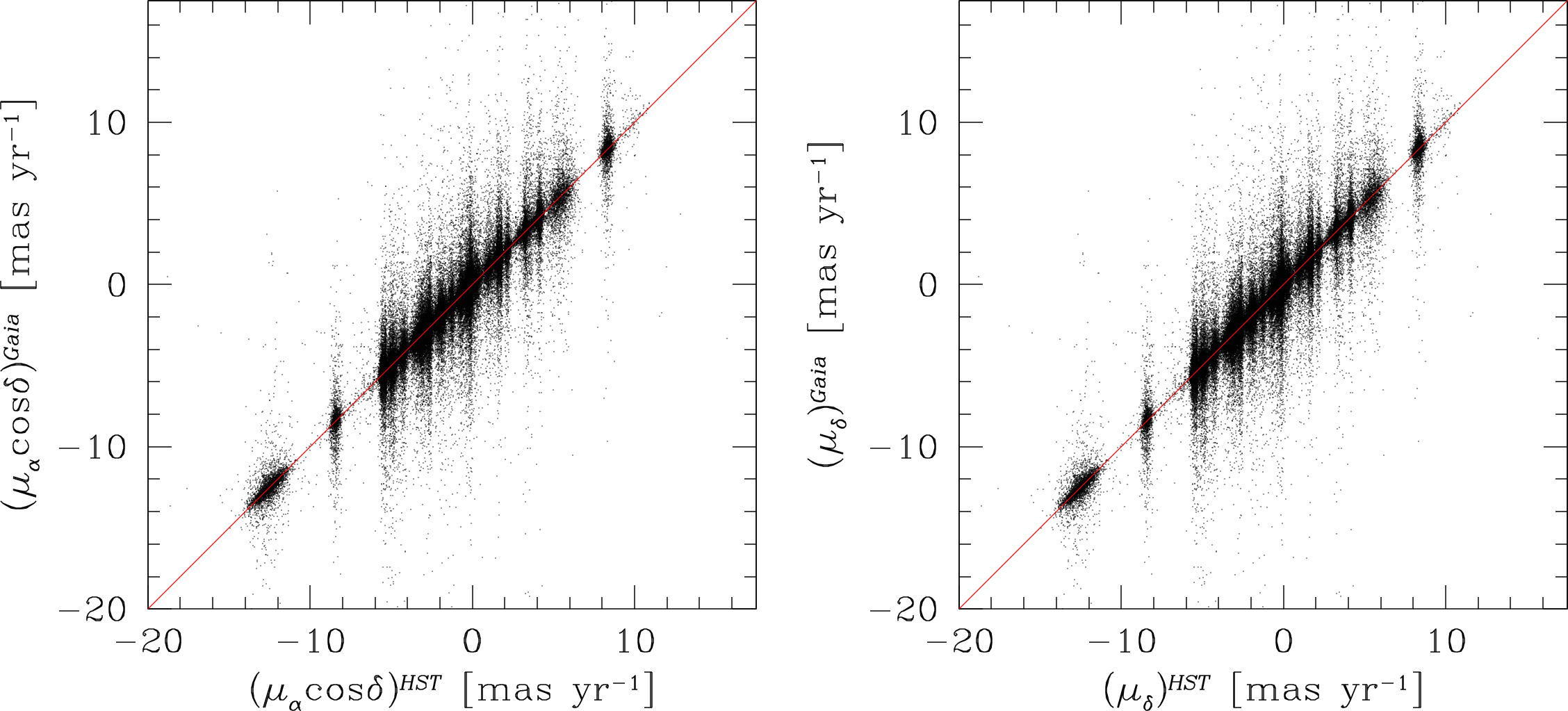}
\caption{Comparison between the \textit{HST} (including the PM zero-points in Table~\ref{tab:abspm}) and \textit{Gaia} PMs. The red lines are the plane bisectors, not a fit to the data.}
\label{fig:abspm}
\end{figure*}

This astrometric registration allows us to directly compare our PMs with those in the \textit{Gaia} EDR3 catalog. We show the result in Fig.~\ref{fig:abspm}. The red lines are the plane bisectors, not a fit to the data. The tight alignment of the points to the plane bisectors shows that our PMs are consistent with those of the \textit{Gaia} EDR3 catalog, and that our absolute registration is accurate. The large scatters along the $y$ directions are due to the poor quality of the \textit{Gaia} PMs in some clusters, likely because of crowding.\looseness=-4
    
\begin{table*}[th]
    \centering
    \caption{PM zero-point needed to transform our relative \textit{HST} PMs onto an absolute reference frame. These values correspond to the absolute PM of each cluster. Uncertainties include both the statistical errors and \textit{Gaia} EDR3 systematic errors.} 
    \label{tab:abspm}
    \begin{tabular}{cc|cc}
    \hline
    \hline
    Cluster & $\Delta (\mu_\alpha \cos\delta, \mu_\delta)$ & Cluster & $\Delta (\mu_\alpha \cos\delta, \mu_\delta)$ \\
    & $[$mas yr$^{-1}$$]$ & & $[$mas yr$^{-1}$$]$ \\
    \hline
	NGC~104  &  $(  5.385  \pm  0.056  ,  -2.340 \pm  0.047)$ & NGC~6352 &  $( -2.168  \pm  0.028  ,  -4.416 \pm  0.028)$ \\
	NGC~288  &  $(  4.154  \pm  0.029  ,  -5.711 \pm  0.028)$ & NGC~6362 &  $( -5.504  \pm  0.027  ,  -4.752 \pm  0.028)$ \\
	NGC~362  &  $(  6.737  \pm  0.039  ,  -2.522 \pm  0.032)$ & NGC~6366 &  $( -0.346  \pm  0.030  ,  -5.176 \pm  0.029)$ \\
	NGC~1261 &  $(  1.577  \pm  0.032  ,  -2.068 \pm  0.035)$ & NGC~6388 &  $( -1.304  \pm  0.029  ,  -2.706 \pm  0.031)$ \\
	NGC~1851 &  $(  2.128  \pm  0.031  ,  -0.646 \pm  0.032)$ & NGC~6397 &  $(  3.231  \pm  0.027  , -17.641 \pm  0.027)$ \\
	NGC~2298 &  $(  3.270  \pm  0.032  ,  -2.165 \pm  0.032)$ & NGC~6441 &  $( -2.571  \pm  0.043  ,  -5.312 \pm  0.040)$ \\
	NGC~2808 &  $(  0.927  \pm  0.054  ,   0.336 \pm  0.098)$ & NGC~6496 &  $( -3.088  \pm  0.037  ,  -9.265 \pm  0.035)$ \\
	NGC~3201 &  $(  8.348  \pm  0.029  ,  -1.965 \pm  0.029)$ & NGC~6535 &  $( -4.220  \pm  0.030  ,  -2.923 \pm  0.033)$ \\
	NGC~4590 &  $( -2.713  \pm  0.029  ,   1.746 \pm  0.030)$ & NGC~6541 &  $(  0.301  \pm  0.030  ,  -8.851 \pm  0.032)$ \\
	NGC~4833 &  $( -8.395  \pm  0.031  ,  -0.952 \pm  0.031)$ & NGC~6584 &  $( -0.111  \pm  0.031  ,  -7.202 \pm  0.028)$ \\
	NGC~5024 &  $( -0.177  \pm  0.033  ,  -1.349 \pm  0.039)$ & NGC~6624 &  $(  0.145  \pm  0.031  ,  -6.944 \pm  0.032)$ \\
	NGC~5053 &  $( -0.333  \pm  0.033  ,  -1.198 \pm  0.034)$ & NGC~6637 &  $( -5.026  \pm  0.031  ,  -5.818 \pm  0.031)$ \\
	NGC~5272 &  $( -0.261  \pm  0.055  ,  -2.674 \pm  0.040)$ & NGC~6652 &  $( -5.491  \pm  0.032  ,  -4.237 \pm  0.029)$ \\
	NGC~5286 &  $(  0.271  \pm  0.044  ,  -0.156 \pm  0.049)$ & NGC~6656 &  $(  9.758  \pm  0.063  ,  -5.665 \pm  0.036)$ \\
	NGC~5466 &  $( -5.371  \pm  0.030  ,  -0.800 \pm  0.031)$ & NGC~6681 &  $(  1.409  \pm  0.030  ,  -4.707 \pm  0.030)$ \\
	NGC~5897 &  $( -5.470  \pm  0.059  ,  -3.405 \pm  0.051)$ & NGC~6715 &  $( -2.691  \pm  0.032  ,  -1.363 \pm  0.030)$ \\
	NGC~5904 &  $(  4.079  \pm  0.036  ,  -9.876 \pm  0.036)$ & NGC~6717 &  $( -3.086  \pm  0.039  ,  -5.003 \pm  0.040)$ \\
	NGC~5927 &  $( -5.058  \pm  0.028  ,  -3.188 \pm  0.028)$ & NGC~6723 &  $(  1.021  \pm  0.033  ,  -2.430 \pm  0.032)$ \\
	NGC~5986 &  $( -4.258  \pm  0.036  ,  -4.569 \pm  0.033)$ & NGC~6752 &  $( -3.155  \pm  0.027  ,  -4.010 \pm  0.028)$ \\
	NGC~6093 &  $( -2.885  \pm  0.052  ,  -5.665 \pm  0.040)$ & NGC~6779 &  $( -1.988  \pm  0.031  ,   1.595 \pm  0.031)$ \\
	NGC~6101 &  $(  1.756  \pm  0.028  ,  -0.245 \pm  0.030)$ & NGC~6791 &  $( -0.421  \pm  0.026  ,  -2.273 \pm  0.026)$ \\
	NGC~6121 &  $(-12.509  \pm  0.028  , -19.012 \pm  0.028)$ & NGC~6809 &  $( -3.434  \pm  0.029  ,  -9.315 \pm  0.028)$ \\
	NGC~6144 &  $( -1.755  \pm  0.031  ,  -2.607 \pm  0.031)$ & NGC~6838 &  $( -3.416  \pm  0.027  ,  -2.655 \pm  0.028)$ \\
	NGC~6171 &  $( -1.932  \pm  0.029  ,  -5.976 \pm  0.028)$ & NGC~6934 &  $( -2.629  \pm  0.035  ,  -4.687 \pm  0.035)$ \\
	NGC~6205 &  $( -3.130  \pm  0.035  ,  -2.505 \pm  0.047)$ & NGC~6981 &  $( -1.231  \pm  0.036  ,  -3.332 \pm  0.031)$ \\
	NGC~6218 &  $( -0.185  \pm  0.028  ,  -6.796 \pm  0.028)$ & NGC~7078 &  $( -0.645  \pm  0.032  ,  -3.803 \pm  0.032)$ \\
	NGC~6254 &  $( -4.766  \pm  0.030  ,  -6.609 \pm  0.028)$ & NGC~7089 &  $(  3.458  \pm  0.139  ,  -2.269 \pm  0.076)$ \\
	NGC~6304 &  $( -4.126  \pm  0.039  ,  -1.004 \pm  0.035)$ & NGC~7099 &  $( -0.718  \pm  0.031  ,  -7.305 \pm  0.030)$ \\
	NGC~6341 &  $( -4.934  \pm  0.031  ,  -0.635 \pm  0.032)$ \\
    \hline
    \end{tabular}
\end{table*}

\section{Description of the publicly available catalogs}\label{cats}

We release the astro-photometric catalogs of the 57 stellar clusters of the GO-13297 program through the MAST archive\footnote{DOI: \protect\dataset[10.17909/jpfd-2m08]{\doi{10.17909/jpfd-2m08}}.\newline See also: \href{https://archive.stsci.edu/hlsp/hacks}{https://archive.stsci.edu/hlsp/hacks}.}. Table~\ref{tab:pmcat} presents the columns of our PM catalogs. For each cluster, we also release the photometric catalogs obtained with the second-pass photometry discussed in Sect.~\ref{datared} for each filter/camera/epoch the data of which were used to compute the PMs. These catalogs will allow users to reproduce the quality selections we applied in Sect.~\ref{kin}.  The description of a typical photometric catalog is provided in Table~\ref{tab:photcat}.\looseness=-4

When using our catalogs, users might notice peculiar features in some CMDs. For example, there are some filter combinations that show a clear, yet nonphysical, split in the SGB and RGB. The sources in these anomalous branches are close to the saturation and present large \texttt{QFIT} and \texttt{RADXS} values, so they can be easily removed from the analysis.\looseness=-4

Our photometric catalogs can contain saturated stars for a given epoch/camera/filter if these sources are not saturated in at least two other epoch/camera/filter combinations so as to enable a PM measurement. As discussed in \citet{2017ApJ...842....6B}, \texttt{KS2} does not deal with saturated pixels, and photometry for saturated stars is instead provided by the first-pass stage of data reduction. Even though the photometric systems of saturated and unsaturated stars were designed to be the same, we have sometimes noticed differences between the regions in the CMDs dominated by objects belonging to either of the groups. These differences can be small zero-points (thus creating a discrete discontinuity in the CMD) or more complex behaviors (like a spread in the CMD or a different RGB slope with respect to that of the unsaturated objects). Caution is again advised when dealing with saturated sources.\looseness=-4

The photometry of the same camera/filter at different epochs is registered on to the same VEGA-mag system. Small zero-point variations can still be present, but they are expected to be small ($\sim$0.01-0.02 mag, i.e., of the order of the uncertainty in the VEGA-mag calibration).\looseness=-4

The astro-photometric catalogs of NGC~362 were made public by \citet{2018ApJ...861...99L}. We include these same catalogs in our online repository, and refer readers to the related paper for their description. For NGC~6352, we provide the astro-photometric catalogs used in \citet{2019ApJ...873..109L}.\looseness=-4

\begin{table*}[th!]
  \caption{Description of a PM catalog.}
  \centering
  \label{tab:pmcat}
  \begin{threeparttable}
  \begin{tabular}{cccc}
    \hline
    \hline
    Column & Name & Unit & Description \\
    \hline
    1  & R.A. & deg & Right Ascension \\
    2  & Dec. & deg & Declination \\
    3  & X & pixel & X master-frame position (in pixel; pixel scale of 40 mas pixel$^{-1}$) \\
    4  & Y & pixel & Y master-frame position (in pixel; pixel scale of 40 mas pixel$^{-1}$) \\ 
    5  & $\mu_\alpha \cos\delta$ & mas yr$^{-1}$ & Corrected PM along $\alpha \cos\delta$ \\
    6  & $\sigma_{\mu_\alpha \cos\delta}$ & mas yr$^{-1}$ & Error on the corrected PM along $\alpha \cos\delta$ \\
    7  & $\mu_\delta$ & mas yr$^{-1}$ &  Corrected PM along $\delta$ \\
    8  & $\sigma_{\mu_\delta}$ & mas yr$^{-1}$ & Error on the corrected PM along $\alpha \cos\delta$ \\
    9  & $\chi^2_{\mu_\alpha \cos\delta}$ & & Reduced $\chi^2$ of the PM fit along $\alpha \cos\delta$ \\
    10 & $\chi^2_{\mu_\delta}$ & & Reduced $\chi^2$ of the PM fit along $\delta$ \\
    11 & $N_{\rm f}^{\rm PM}$ & & Number of exposures initially considered in the PM fit \\
    12 & $N_{\rm u}^{\rm PM}$ & & Number of exposures actually used in the PM fit \\
    13 & $\Delta {\rm time}$ & yr & Temporal baseline of the PM fit \\
    14 & $(\mu_\alpha \cos\delta)_{\rm raw}$ & mas yr$^{-1}$ & Raw PM along $\alpha \cos\delta$ \\
    15 & $(\sigma_{\mu_\alpha \cos\delta})_{\rm raw}$ & mas yr$^{-1}$ & Error on the raw PM along $\alpha \cos\delta$ \\
    16 & $(\mu_\delta)_{\rm raw}$ & mas yr$^{-1}$ & Raw PM along $\delta$ \\
    17 & $(\sigma_{\mu_\delta})_{\rm raw}$ & mas yr$^{-1}$ & Error on the raw PM along $\delta$ \\
    18 & Corr\_Flag & & Flags that tells if the PM is a-posteriori corrected for systematics  \\
    19 & ID & & ID number of the source \\
    \hline
  \end{tabular}
  \begin{tablenotes}[flushleft]
  \item \textbf{Note.} (i) The ID number of a source is the same in all catalogs for the same cluster. (ii) (X,Y) positions are defined at a specific epoch about halfway between the first and last epochs used in the PM computation. This reference epoch is provided in the header of the file. (iii) ``Corr\_Flag" is set to 1 if the PM was a-posteriori corrected for high-frequency systematics, and to 0 if otherwise.
  \end{tablenotes}
  \end{threeparttable}
\end{table*}

\begin{table*}[th!]
  \caption{Description of a photometric catalog for one filter/camera/epoch.}
  \centering
  \label{tab:photcat}
  \begin{threeparttable}
  \begin{tabular}{ccc}
    \hline
    \hline
    Column & Name & Description \\
    \hline
    1  & $m$ & Calibrated VEGA magnitude \\
    2  & $\sigma_m$ & Photometric rms \\
    3  & \texttt{QFIT} & Quality-of-PSF-fit (QFIT) parameter \\
    4  & $o$ & Fractional flux within the fitting radius prior to neighbor subtraction \\
    5  & $N_{\rm f}^{\rm phot}$ & Number of exposures in which a source was found \\
    6  & $N_{\rm u}^{\rm phot}$ & Number of exposures used to measure the flux of a source \\
    7  & \texttt{RADXS} & Excess/deficiency of flux outside the core of the star \\
    8  & sky & Sky in electrons \\
    9  & $\sigma_{\rm sky}$ & Sky rms in electrons \\
    10 & sat & Saturation flag \\
    11 & ID & ID number of the source \\
    \hline
  \end{tabular}
  \begin{tablenotes}[flushleft]
  \item \textbf{Note.} (i): The ID number of a source is the same in all catalogs for the same cluster. (ii): All the values for a source are set to 0 if it is not measured in this filter/camera/epoch (iii) All the values for a source except the calibrated magnitude are set to 0 if it is saturated in this catalog. (iv): The saturation flag is set to 9 if the star is saturated. If multiple exposure times are present in the data set, the saturation flag is 0 for a measurement obtained from the longest exposure(s), and progressively increases (by 1) with the decreasing exposure time of the image(s) from which it was measured. (v) To estimate the significance of a source over the sky, first convert the calibrated VEGA magnitudes into instrumental fluxes (in units of electrons): flux $=$ $10^{-0.4({\rm mag} - {\rm VEGA\_zp})}$ The VEGA-magnitude zero-point is provided in the header of the catalog.
  \end{tablenotes}
  \end{threeparttable}
\end{table*}

\section{Kinematic profiles and table}\label{app:kin}

\begin{figure*}[th!]  \centering \includegraphics[width=\textwidth]{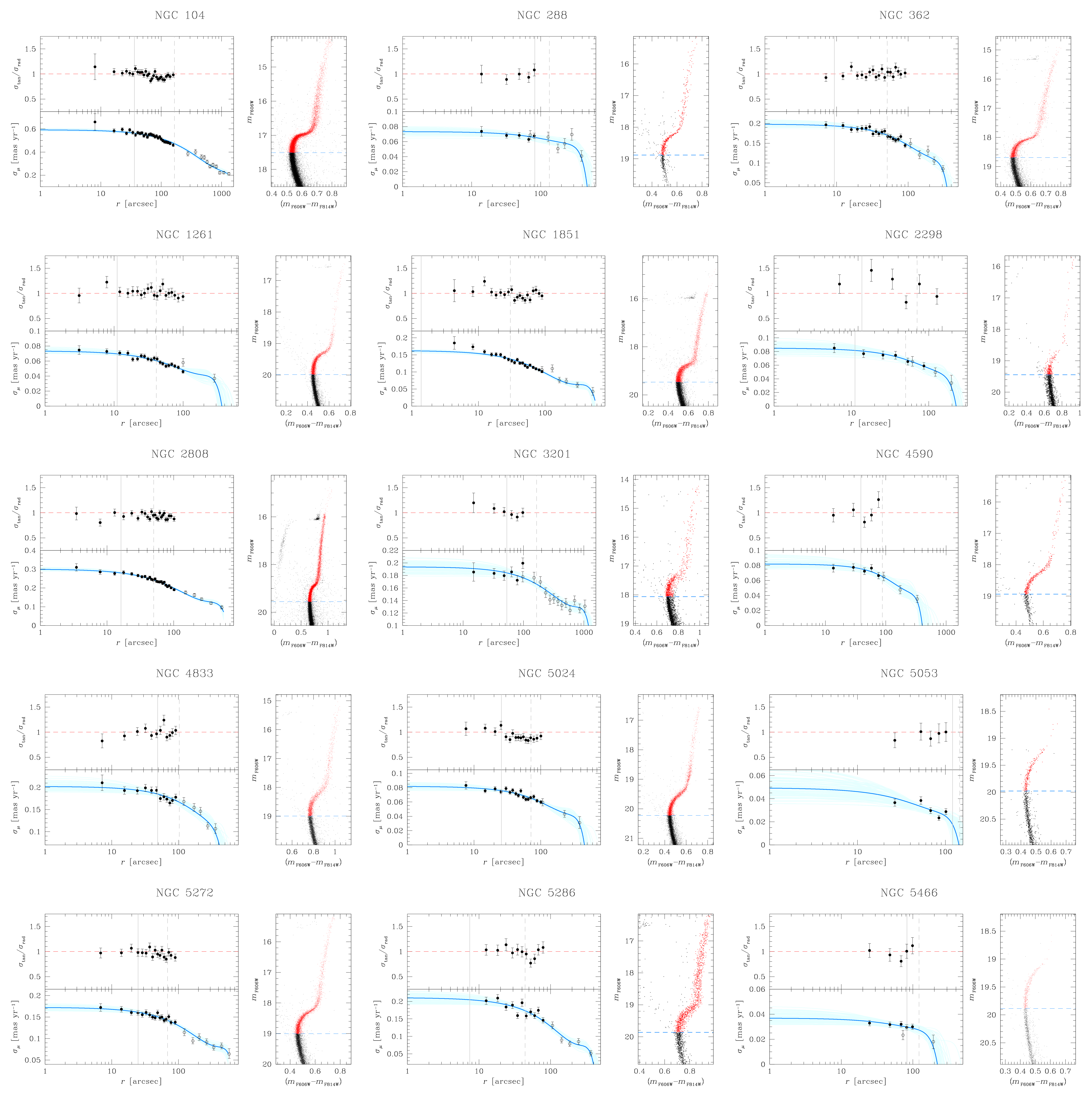} \caption{Velocity-dispersion and anisotropy radial profiles for NGC~104, NGC~288, NGC~362, NGC~1251, NGC~1851, NGC~2298, NGC~2808, NGC~3201, NGC~4590, NGC~4833, NGC~5024, NGC~5053, NGC~5272, NGC~5286 and NGC~5466. For each cluster, an $m_{\rm F606W}$ versus $(m_{\rm F606W}-m_{\rm F814W})$ CMD is shown. Red points are stars used for the kinematic analysis, while black dots represent all other objects. The azure, dashed, horizontal line is set at the MS turnoff level. The velocity dispersion $\sigma_\mu$ as a function of distance from the center of the cluster (in arcsec) is plotted in the bottom-left panel. Black, filled points are obtained by means of the \textit{HST} PMs; black open circles are the \textit{Gaia}-EDR3 measurements of \citet{2021MNRAS.505.5978V}. The blue line is 4-th-order polynomial fit to the data. The cyan lines show 100 random solutions of the polynomial fit. In the top-left panel, we finally show the anisotropy ($\sigma_{\rm tan}/\sigma_{\rm rad}$) radial profile. Only \textit{HST} data is shown because the catalogs \citet{2021MNRAS.505.5978V} do not contain this piece of information. The red, dashed, horizontal line is set to 1 (isotropic velocity distribution). The gray lines in the left panels are set at the $r_{\rm c}$ (solid) and $r_{\rm h}$ (dashed) radii of the cluster. These lines are shown only if they are within the boundaries of the plot.}
\label{fig:kin1}
\end{figure*}

\begin{figure*}
\centering
\includegraphics[width=\textwidth]{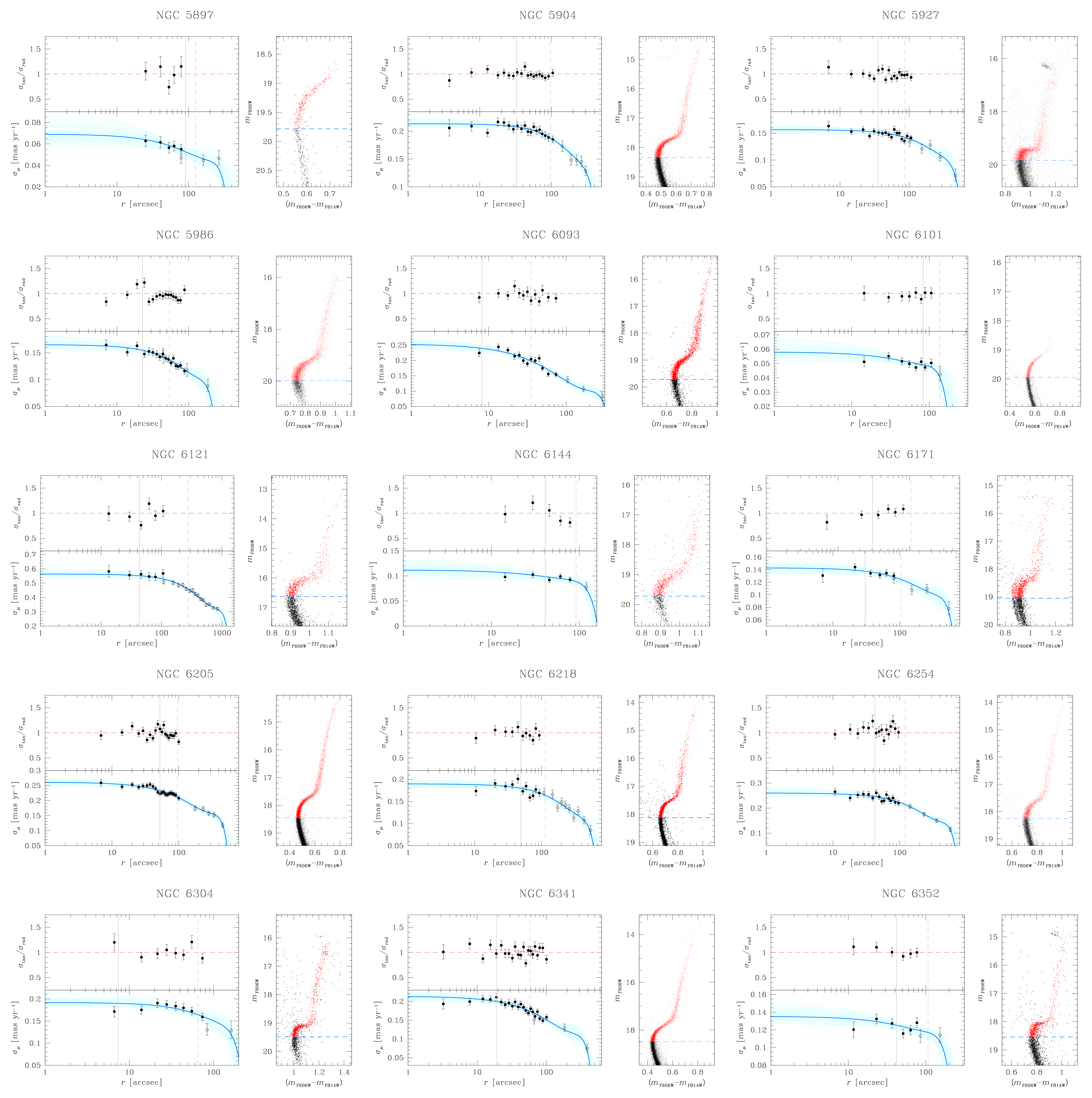}
\caption{Similar to Fig.~\ref{fig:kin1}, but for NGC~5897, NGC~5904, NGC~5927, NGC~5986, NGC~6093, NGC~6101, NGC~6121, NGC~6144, NGC~6171, NGC~6205, NGC~6218, NGC~6254, NGC~6304, NGC~6341, and NGC~6352. The CMD of NGC~5897 is cut at the SGB level because only one exposure mapping the RGB is available in the GO-10775 data. Having only two epochs at our disposal for this GC, the PM fit for these bright stars would have been forced to pass through the only first-epoch point, thus making the PM measurement uncertain.}
\label{fig:kin2}
\end{figure*}

\begin{figure*}
\centering
\includegraphics[width=\textwidth]{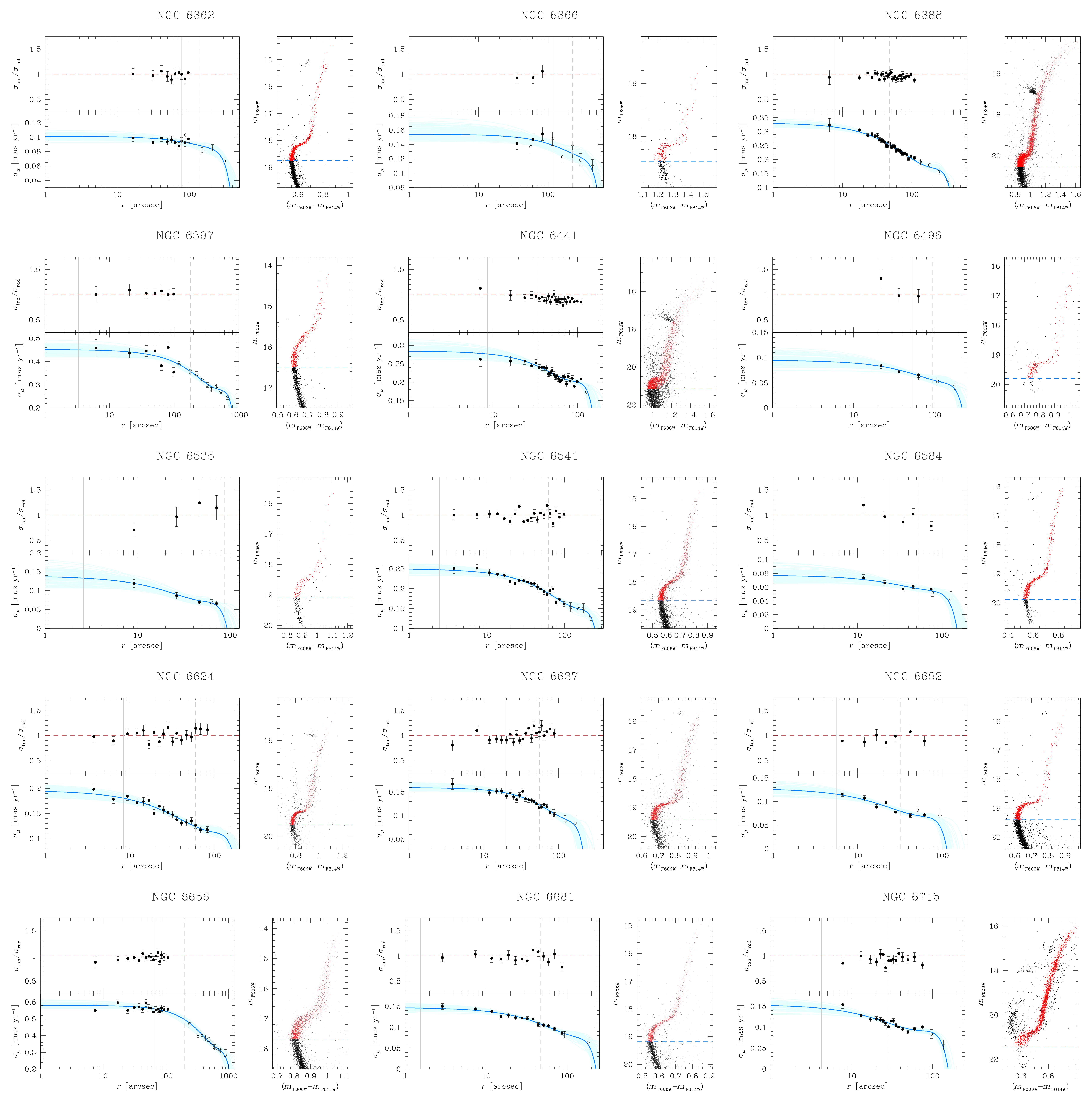}
\caption{Similar to Figs.~\ref{fig:kin1} and \ref{fig:kin2}, but for NGC~6362, NGC~6366, GC~6388, NGC~6397, NGC~6441, NGC~6496, NGC~6535, NGC~6541, NGC~6584, NGC~6624, NGC~6637, NGC~6652, NGC~6656, NGC~6681 and NGC~6715.}
\label{fig:kin3}
\end{figure*}

\begin{figure*}
\centering
\includegraphics[width=\textwidth]{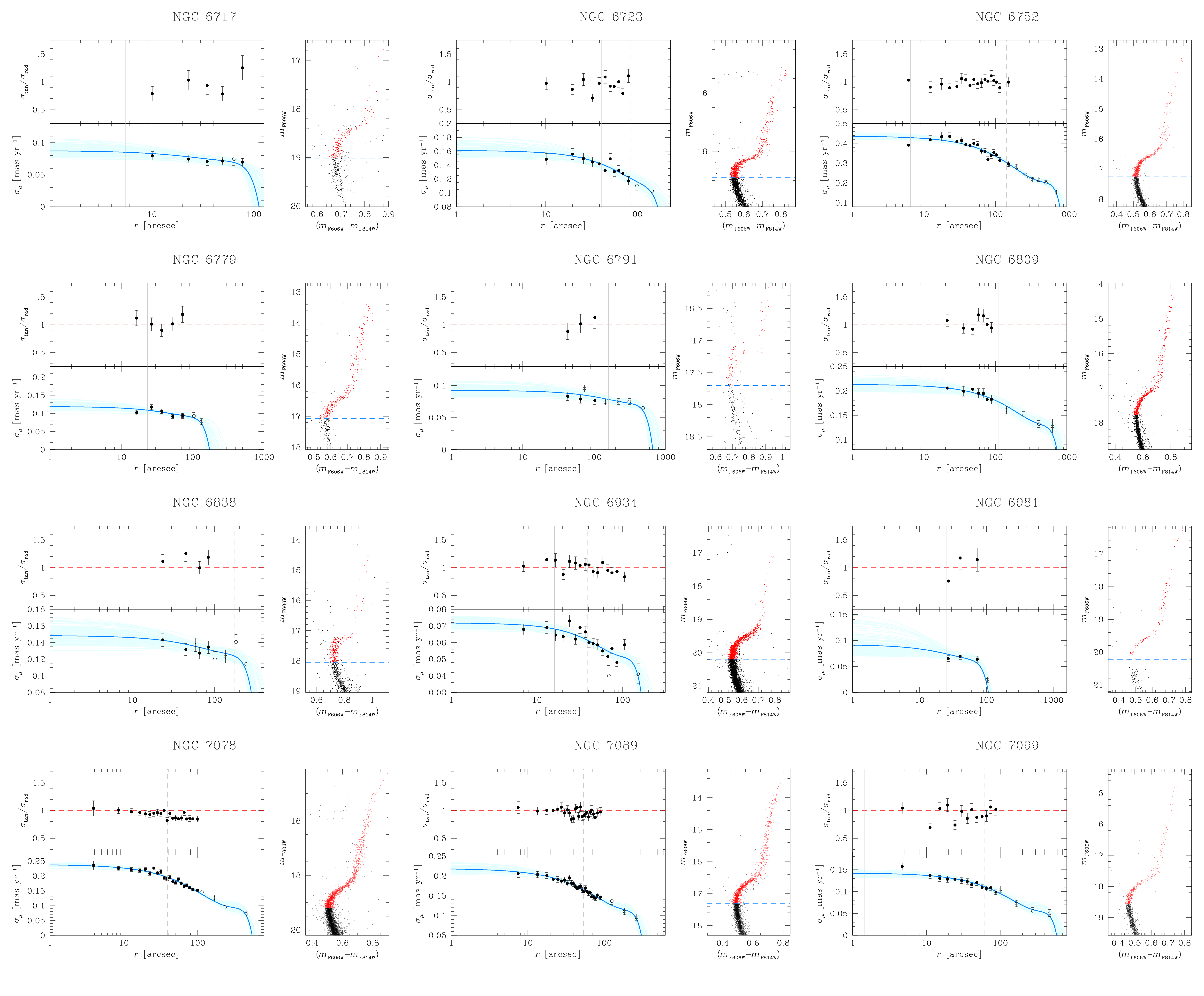}
\caption{Similar to Figs.~\ref{fig:kin1}, \ref{fig:kin2} and \ref{fig:kin3}, but for NGC~6717, NGC~6723, NGC~6752, NGC~6779, NGC~6791, NGC~6809, NGC~6838, NGC~6934, NGC~6981, NGC~7078, NGC~7078 and NGC~7099.}
\label{fig:kin4}
\end{figure*}

In this Section, we present velocity-dispersion and anisotropy radial profiles. We refer to Sect.~\ref{kin} for the detailed description of Figs.~\ref{fig:kin1}, \ref{fig:kin2}, \ref{fig:kin3}, and \ref{fig:kin4}. Table~\ref{tab:vdisp} provides the values of $\sigma_\mu$ at $r = 0$ arcsec, $r_{\rm c}$ and $r_{\rm h}$. The individual measurements are available on our website.\looseness=-4

\begin{table*}[th]
    \centering
    \caption{Velocity dispersions at $r=0$ arcsec, $r_{\rm c}$ and $r_{\rm h}$ for all clusters. Not enough data are available to obtain a reliable estimate of the velocity dispersion at large radii for NGC~5053, NGC~6535 and NGC~6717.\\}
    \label{tab:vdisp}
    \begin{tabular}{cccc|cccc}
    \hline
    \hline
    Cluster & $\sigma_\mu^{r=0}$ & $\sigma_\mu^{r=r_{\rm c}}$ & $\sigma_\mu^{r=r_{\rm h}}$ & Cluster & $\sigma_\mu^{r=0}$ & $\sigma_\mu^{r=r_{\rm c}}$ & $\sigma_\mu^{r=r_{\rm h}}$\\
    & $[$mas yr$^{-1}$$]$ & $[$mas yr$^{-1}$$]$ & $[$mas yr$^{-1}$$]$ & & $[$mas yr$^{-1}$$]$ & $[$mas yr$^{-1}$$]$ & $[$mas yr$^{-1}$$]$ \\
    \hline
    NGC~0104 & $ 0.592 \pm 0.005 $ & $ 0.563 \pm 0.004 $ & $ 0.473 \pm 0.004 $ & NGC~6352 & $ 0.135 \pm 0.006 $ & $ 0.125 \pm 0.002 $ & $ 0.117 \pm 0.003 $ \\
    NGC~0288 & $ 0.073 \pm 0.003 $ & $ 0.065 \pm 0.002 $ & $ 0.061 \pm 0.002 $ & NGC~6362 & $ 0.101 \pm 0.003 $ & $ 0.093 \pm 0.002 $ & $ 0.088 \pm 0.002 $ \\
    NGC~0362 & $ 0.198 \pm 0.005 $ & $ 0.194 \pm 0.003 $ & $ 0.171 \pm 0.002 $ & NGC~6366 & $ 0.154 \pm 0.007 $ & $ 0.137 \pm 0.003 $ & $ 0.126 \pm 0.005 $ \\
    NGC~1261 & $ 0.073 \pm 0.002 $ & $ 0.070 \pm 0.001 $ & $ 0.061 \pm 0.001 $ & NGC~6388 & $ 0.331 \pm 0.006 $ & $ 0.316 \pm 0.005 $ & $ 0.254 \pm 0.001 $ \\
    NGC~1851 & $ 0.163 \pm 0.003 $ & $ 0.162 \pm 0.002 $ & $ 0.138 \pm 0.001 $ & NGC~6397 & $ 0.452 \pm 0.016 $ & $ 0.450 \pm 0.015 $ & $ 0.356 \pm 0.006 $ \\
    NGC~2298 & $ 0.085 \pm 0.005 $ & $ 0.081 \pm 0.003 $ & $ 0.067 \pm 0.002 $ & NGC~6441 & $ 0.285 \pm 0.012 $ & $ 0.277 \pm 0.008 $ & $ 0.242 \pm 0.003 $ \\
    NGC~2808 & $ 0.300 \pm 0.004 $ & $ 0.280 \pm 0.003 $ & $ 0.243 \pm 0.001 $ & NGC~6496 & $ 0.095 \pm 0.009 $ & $ 0.067 \pm 0.003 $ & $ 0.056 \pm 0.004 $ \\
    NGC~3201 & $ 0.194 \pm 0.005 $ & $ 0.184 \pm 0.003 $ & $ 0.166 \pm 0.003 $ & NGC~6535 & $ 0.139 \pm 0.017 $ & $ 0.133 \pm 0.014 $ &          -          \\
    NGC~4590 & $ 0.082 \pm 0.004 $ & $ 0.075 \pm 0.002 $ & $ 0.065 \pm 0.002 $ & NGC~6541 & $ 0.250 \pm 0.006 $ & $ 0.247 \pm 0.006 $ & $ 0.188 \pm 0.002 $ \\
    NGC~4833 & $ 0.202 \pm 0.007 $ & $ 0.185 \pm 0.002 $ & $ 0.166 \pm 0.003 $ & NGC~6584 & $ 0.078 \pm 0.006 $ & $ 0.066 \pm 0.002 $ & $ 0.058 \pm 0.002 $ \\
    NGC~5024 & $ 0.082 \pm 0.002 $ & $ 0.076 \pm 0.001 $ & $ 0.065 \pm 0.001 $ & NGC~6624 & $ 0.196 \pm 0.006 $ & $ 0.181 \pm 0.004 $ & $ 0.125 \pm 0.002 $ \\
    NGC~5053 & $ 0.049 \pm 0.007 $ &          -          &          -          & NGC~6637 & $ 0.160 \pm 0.005 $ & $ 0.148 \pm 0.002 $ & $ 0.122 \pm 0.002 $ \\
    NGC~5272 & $ 0.173 \pm 0.004 $ & $ 0.161 \pm 0.002 $ & $ 0.142 \pm 0.002 $ & NGC~6652 & $ 0.128 \pm 0.007 $ & $ 0.116 \pm 0.004 $ & $ 0.080 \pm 0.002 $ \\
    NGC~5286 & $ 0.211 \pm 0.009 $ & $ 0.205 \pm 0.007 $ & $ 0.173 \pm 0.003 $ & NGC~6656 & $ 0.581 \pm 0.010 $ & $ 0.560 \pm 0.005 $ & $ 0.496 \pm 0.008 $ \\
    NGC~5466 & $ 0.037 \pm 0.003 $ & $ 0.029 \pm 0.001 $ & $ 0.028 \pm 0.002 $ & NGC~6681 & $ 0.147 \pm 0.004 $ & $ 0.146 \pm 0.004 $ & $ 0.108 \pm 0.002 $ \\
    NGC~5897 & $ 0.069 \pm 0.006 $ & $ 0.052 \pm 0.002 $ & $ 0.048 \pm 0.003 $ & NGC~6715 & $ 0.154 \pm 0.007 $ & $ 0.145 \pm 0.005 $ & $ 0.110 \pm 0.001 $ \\
    NGC~5904 & $ 0.213 \pm 0.003 $ & $ 0.207 \pm 0.002 $ & $ 0.187 \pm 0.003 $ & NGC~6717 & $ 0.088 \pm 0.007 $ & $ 0.085 \pm 0.006 $ &          -          \\
    NGC~5927 & $ 0.157 \pm 0.003 $ & $ 0.152 \pm 0.002 $ & $ 0.142 \pm 0.002 $ & NGC~6723 & $ 0.161 \pm 0.007 $ & $ 0.140 \pm 0.003 $ & $ 0.120 \pm 0.003 $ \\
    NGC~5986 & $ 0.159 \pm 0.005 $ & $ 0.149 \pm 0.002 $ & $ 0.134 \pm 0.001 $ & NGC~6752 & $ 0.436 \pm 0.009 $ & $ 0.429 \pm 0.008 $ & $ 0.302 \pm 0.005 $ \\
    NGC~6093 & $ 0.254 \pm 0.008 $ & $ 0.239 \pm 0.006 $ & $ 0.199 \pm 0.003 $ & NGC~6779 & $ 0.119 \pm 0.008 $ & $ 0.108 \pm 0.004 $ & $ 0.096 \pm 0.003 $ \\
    NGC~6101 & $ 0.058 \pm 0.003 $ & $ 0.049 \pm 0.001 $ & $ 0.041 \pm 0.004 $ & NGC~6791 & $ 0.093 \pm 0.006 $ & $ 0.078 \pm 0.002 $ & $ 0.074 \pm 0.003 $ \\
    NGC~6121 & $ 0.563 \pm 0.015 $ & $ 0.552 \pm 0.010 $ & $ 0.456 \pm 0.004 $ & NGC~6809 & $ 0.213 \pm 0.008 $ & $ 0.177 \pm 0.004 $ & $ 0.161 \pm 0.004 $ \\
    NGC~6144 & $ 0.112 \pm 0.008 $ & $ 0.098 \pm 0.003 $ & $ 0.090 \pm 0.003 $ & NGC~6838 & $ 0.149 \pm 0.008 $ & $ 0.132 \pm 0.003 $ & $ 0.125 \pm 0.004 $ \\
    NGC~6171 & $ 0.143 \pm 0.005 $ & $ 0.137 \pm 0.003 $ & $ 0.124 \pm 0.003 $ & NGC~6934 & $ 0.072 \pm 0.002 $ & $ 0.068 \pm 0.001 $ & $ 0.061 \pm 0.001 $ \\
    NGC~6205 & $ 0.261 \pm 0.006 $ & $ 0.233 \pm 0.002 $ & $ 0.211 \pm 0.002 $ & NGC~6981 & $ 0.092 \pm 0.016 $ & $ 0.073 \pm 0.004 $ & $ 0.065 \pm 0.004 $ \\
    NGC~6218 & $ 0.190 \pm 0.005 $ & $ 0.180 \pm 0.003 $ & $ 0.163 \pm 0.003 $ & NGC~7078 & $ 0.238 \pm 0.003 $ & $ 0.237 \pm 0.003 $ & $ 0.196 \pm 0.001 $ \\
    NGC~6254 & $ 0.260 \pm 0.007 $ & $ 0.245 \pm 0.003 $ & $ 0.214 \pm 0.004 $ & NGC~7089 & $ 0.219 \pm 0.006 $ & $ 0.204 \pm 0.003 $ & $ 0.166 \pm 0.002 $ \\
    NGC~6304 & $ 0.192 \pm 0.006 $ & $ 0.190 \pm 0.005 $ & $ 0.162 \pm 0.005 $ & NGC~7099 & $ 0.143 \pm 0.004 $ & $ 0.142 \pm 0.004 $ & $ 0.111 \pm 0.002 $ \\
    NGC~6341 & $ 0.213 \pm 0.005 $ & $ 0.200 \pm 0.003 $ & $ 0.174 \pm 0.002 $ \\
    \hline
    \end{tabular}
\end{table*}

\bibliography{GC_PMs}{}
\bibliographystyle{aasjournal}

\end{document}